\documentclass[twocolumn,showpacs,preprintnumbers,amsmath,amssymb,prd,letterpaper,floatfix,nofootinbib,superscriptaddress]{revtex4-1}
\usepackage{graphics}
\usepackage{graphicx}
\usepackage{epsfig}
\usepackage[usenames]{color}
\usepackage{longtable}
\PassOptionsToPackage{hyphens}{url}\usepackage{hyperref}
\usepackage{breakurl}
\usepackage{latexsym}
\usepackage{amsmath}
\usepackage{amsthm}
\usepackage{amsfonts}
\usepackage{amssymb}
\usepackage{dsfont}
\usepackage{bm}

\newcommand\T{\rule{0pt}{2.6ex}}
\newcommand\B{\rule[-1.2ex]{0pt}{0pt}}

\begin{document}


\title{$D$ $\pi$ scattering and $D$ meson resonances from lattice QCD}



\author{Daniel Mohler}
\email{dmohler@fnal.gov}
\affiliation{TRIUMF, 4004 Wesbrook Mall Vancouver, BC V6T 2A3, Canada}

\author{Sasa Prelovsek}
\email{sasa.prelovsek@ijs.si}
\affiliation{Department of Physics, University of Ljubljana, Slovenia}
\affiliation{Jozef Stefan Institute, Ljubljana, Slovenia}

\author{R.~M.~Woloshyn}
\email{rwww@triumf.ca}
\affiliation{TRIUMF, 4004 Wesbrook Mall Vancouver, BC V6T 2A3, Canada}

\date{\today}

\begin{abstract}
A first exploratory lattice QCD simulation is presented aimed at extracting the masses and widths of the broad scalar $D_0^*(2400)$ and the axial $D_1(2430)$ charm-light resonances. For that purpose $D\pi$ and $D^*\pi$ scattering are simulated, and  the resonance parameters are extracted using a Breit-Wigner fit of the resulting phase shifts. We use a single two-flavor dynamical  ensemble with   $m_\pi\approx 266~$MeV,  $a\simeq 0.124~$fm and a rather small volume $V=16^3\times 32$. The resulting $D_0^*(2400)$  mass is $351\pm 21~$MeV above the spin-average $\tfrac{1}{4}(m_D+3m_{D^*})$, in agreement with the experimental value of $347\pm 29~$MeV above. The resulting $D_0^*\to D\pi$ coupling $g^{lat}=2.55\pm 0.21~$GeV is close to the experimental value $g^{exp}\le1.92\pm 0.14~$GeV, where $g$ parametrizes the width $\Gamma\equiv g^2p^*/s$. The resonance parameters for the broad $D_1(2430)$
are also found close to the experimental values; these are obtained by appealing to the heavy quark limit, where the neighboring resonance $D_1(2420)$ is narrow. The calculated $I=1/2$ scattering lengths are $a_0=0.81~\pm 0.14~$fm for $D\pi$ and $a_0=0.81\pm 0.17~$fm for $D^*\pi$ scattering. The simulation of the scattering in these channels incorporates quark-antiquark as well as multi-hadron interpolators, and the distillation method is used for contractions. In addition,   
the ground and several excited charm-light and charmonium states with various $J^P$ are calculated using standard quark-antiquark interpolators.
\end{abstract}


\maketitle

\section{Introduction}

In the spectrum of $D$ mesons, the six lowest states are well established experimentally \cite{pdg12,Abe:2003zm,Abe:2004sm,Link:2003bd,Aubert:2009wg}. These correspond to $1S$ and  $1P$ states of $u\bar c$ within the quark model. The knowledge of the higher radial and orbital excitations is poor, with the only experimental information based on a BaBar study in 2010 \cite{delAmoSanchez:2010vq}, which found several new resonances whose quantum numbers are mostly unknown and the states are unconfirmed by any other experiment \cite{pdg12}. 

The observed  pattern of masses and widths for the six lightest $D$ mesons can be understood qualitatively by assuming $u\bar c$ valence structure and by appealing to the $m_c\to \infty$ limit \cite{PhysRevLett.66.1130}. The masses and widths are then independent of heavy quark spin $\vec s_c$,  and the total angular momentum of the light quark $\vec j_q=\vec s_q+\vec L$ is a good quantum number\footnote{We use small $j$ when referring to heavy quark limit, while $J$ is the total spin 
in the heavy quark limit or away from it. }, while the total angular momentum of the state is $J^P=j^P\pm s_c=j^P\pm \tfrac{1}{2}$.  The $D$ and $D^*$ belong to the $1S$ heavy quark doublet with $j^P=\tfrac{1}{2}^-$    and $J^P=(0^-,1^-)$. The observed narrow $1P$ states $D_1(2420)$ and $D_2^*(2460)$ are candidates for $j^P=\tfrac{3}{2}^+$ with  $J^P=(1^+,2^+)$, since only D-wave decays are allowed in the heavy-quark limit, making them naturally narrow \cite{PhysRevLett.66.1130}. On the other hand, the $j^P=\tfrac{1}{2}^+$ states with  $J^P=(0^+,1^+)$ decay via S-wave decays in this limit \cite{PhysRevLett.66.1130} and are related to the broad resonances $D_0^*(2400)$ and $D_1(2430)$.  

The excited $D$-mesons are particularly interesting in view of several persisting puzzles related to the semileptonic $B\to D^{**} l\bar \nu_l$, where  $D^{**}$ collectively indicates all $D$-mesons, except for the ground state $j^{P}=\tfrac{1}{2}^-$ doublet (see for example \cite{Bigi:2007qp,Uraltsev:2004ta,LeYaouanc:2000cj,Blossier:2009vy}). While experiments indicate that $B\to D^{**}_{j^P=1/2^+}l\nu_l$ are more likely, theory strongly favors  $B\to D^{**}_{j^P=3/2^+}l\nu_l$. A second puzzle is related to the fact that the exclusive modes into the known charm hadrons do not saturate the inclusive $B\to X_c l\bar \nu$.

$D$-meson spectroscopy can be addressed quantitatively using lattice QCD. In recent  studies\cite{Mohler:2011ke,Namekawa:2011wt,Bali:2011dc} masses for $D$ (and $D_{s}$) mesons were calculated in lattice QCD with dynamical quarks close to physical values. In these calculations the mesons were interpolated by standard quark-antiquark operators. This may be problematic for the broad states $D_0^*(2400)$ and $D_1(2430)$ which were not very well described by the simulation in \cite{Mohler:2011ke}. As well, in the charm-strange sector the scalar $D_{s0}^*(2317)$ and the axial $D_{s1}(2460)$ were discovered below $DK$ and $D^*K$ thresholds, which is significantly lower than anticipated. The closeness of the masses for the scalar states $M_{D_0^*}=2318\pm 29~$MeV and $M_{D_{s0}}=2317.8\pm 0.6~$MeV \cite{pdg12} is not natural within a picture where the mass is dominated by the valence quark-antiquark content. This has led to suggestions that nearby thresholds may play an important role for the $D_{s0}^*(2317)$ state (see, for example, \cite{Geng:2010vw}). For these reasons a natural next step for lattice calculations of charm mesons is the explicit inclusion of multi-hadron operators and the treatment of states with open hadronic decay channels as resonances. In this work we begin with a study of $D$ mesons where, in a lattice simulation, the relevant S-wave $D\pi$ and $D^*\pi$ decay channels of scalar and axial mesons are open over a large range of heavier than physical pion mass. In contrast, a corresponding study
in the $D_{s}$ sector would likely require a simulation tuned to near physical quark masses to achieve proximity of $D_{s0}^*$ and $D_{s1}$ with the $DK$ and $D^*K$ thresholds \cite{Mohler:2011ke}.
 
 Calculations are carried out using a lattice simulation with dynamical $u,d$ quarks. Correlation functions are constructed with quark-antiquark interpolators and, for the first time, also with $D\pi$ or $D^*\pi$ interpolators to study the relevant scattering channels. Our aim is to describe the two observed broad states $D_0^*(2400)$ and $D_1(2430)$ as resonances, so we simulate $D\pi$ and $D^*\pi$ scattering and extract the corresponding phase shifts for the first time. The S-wave phase shift for $D\pi$ scattering is extracted using L\" uscher's formula and a Breit-Wigner fit of the phase shift renders the $D_0^*(2400)$ resonance mass and width.

An analogous procedure is used for the two $D_1$ resonances in $D^*\pi$ scattering with $J^P=1^+$, but in this case S-wave and D-wave are present  and our  analysis relies on the following model assumptions: (i) we appeal to the heavy quark limit \cite{PhysRevLett.66.1130}, where the broad $D_1(2430)$ appears only in S-wave and the narrow $D_1(2420)$ appears only in D-wave; (ii) one energy level is associated with the narrow D-wave resonance $D_1(2420)$ and its mass is extracted; (iii) we  assume that the contribution of the D-wave phase shift to the other three energy levels is negligible which is a valid assumption in the limit of a very narrow resonance. Using a Breit-Wigner fit of the resulting $D^*\pi$ S-wave phase shift, we extract the mass and the width of the broad $D_1(2430)$.  

The remaining three members ($D$, $D^*$ and $D_2(2460)$)  of the $1S$ and $1P$ multiplets are stable or very narrow in our lattice simulation with $m_\pi\approx 266$ MeV. We equate the masses of these states directly to the quark-antiquark energy levels on the lattice, as in all lattice simulations up to now. In addition, the masses of the ground and  excited states in channels with $J^P=0^-,1^-,2^\pm$ are extracted. Some of these correspond to the still poorly known orbital and radial $D$ meson excitations.  

The $D\pi$ scattering in the $I=1/2$ channel has been addressed on the lattice only indirectly by simulating the scalar semileptonic $D\to \pi$ form factor $f_0$ \cite{Flynn:2007ki}.  
Various scattering channels in the charm sector were simulated in \cite{Liu:2008rza}, but the attractive $I=1/2$ scattering $D\pi$ or $D^{*}\pi$ has not been directly simulated yet. 
While the scattering lengths can not be measured, their calculation is of theoretical interest and we calculate the $D\pi$ and $D^*\pi$ scattering lengths on the lattice which can be compared to other types of calculations \cite{PhysRevLett.17.616,Liu:2009uz,Geng:2010vw,Flynn:2007ki,Guo:2009ct,Liu:2011mi}.

The present study  of charm-light spectroscopy  requires good control over heavy-quark discretization effects, as for example provided by the Fermilab method \cite{ElKhadra:1996mp}. In \cite{Mohler:2011ke} the spectrum of low-lying charmonium states was used to validate the approach. Motivated by these results on the low-lying charmonium spectrum a large number of non-exotic charmonium states up to spin $3$ are studied in the present work.\\ 

The present paper is organized as follows. 
Section \ref{calc} outlines the calculational setup. Details about the gauge configurations, the calculation of quark propagators and the determination of the charm quark hopping parameter $\kappa_c$ are discussed. Section \ref{charmonium} presents results for the spectrum of low-lying charmonium states. Encouraged by these results, we simulate  $D\pi$ and $D^*\pi$ scattering in Section \ref{dmesons} and extract information on  scalar and axial resonances. For completeness some results with regular quark-antiquark ($q\bar{q}$) interpolators in other $J^P$ channels are presented. Section \ref{outlook} contains a summary and discussion. Tables of lattice interpolating fields as well as details about fits and fit results are included in the appendix.
 
\section{\label{calc}Calculational setup}

Gauge field configurations were generated with $n_f=$2 flavors of tree level improved Wilson-Clover fermions \cite{Hasenfratz:2008ce,Hasenfratz:2008fg}. The gauge links in the action have been smeared using normalized hypercubic (nHYP) smearing \cite{Hasenfratz:2007rf} with parameters $(\alpha_1,\alpha_2,\alpha_3)=(0.75,0.6,0.3)$. In these simulations the gauge fields have been generated with periodic boundary conditions and the fermion fields obey periodic boundary conditions in space and anti-periodic boundary conditions in time. The same configurations were used previously in a coupled channel analysis of the $\rho$ meson \cite{Lang:2011mn} and in a study of $K\pi$-scattering \cite{Lang:2012sv}. Table \ref{gauge_configs} lists some further details about the gauge configurations. For the determination of the lattice spacing $a$ and the strange quark hopping parameter $\kappa_s$ please refer to \cite{Lang:2011mn} and  \cite{Lang:2012sv}, respectively. 

\begin{table}[t]
\begin{ruledtabular}
\begin{tabular}{ccccccc}
$N_L^3\times N_T$ & $\kappa_l$ & $\beta$ & $a$[fm] & $L$[fm] & \#configs & $m_\pi$[MeV]\\ 
\hline
$16^3\times32$ & 0.1283 & 7.1 & 0.1239(13) & 1.98 & 280/279 & 266(3)(3) \\
\end{tabular}
\end{ruledtabular}
\caption{\label{gauge_configs}Details of $N_f=2$ gauge configurations: $N_L$
and $N_T$ denote the number of lattice points in spatial and time directions. The first error on $m_\pi$ is statistical while the second error is
from the determination of the lattice scale. Observables are based on 279 or 280 configurations. For details see \cite{Lang:2011mn}.}
\end{table}

To calculate the quark propagation the \texttt{dfl\_sap\_gcr} inverter from L\"uscher's DD-HMC package \cite{Luscher:2007se,Luscher:2007es} is used for the light and strange quarks and the same inverter without low mode deflation is used for the charm quarks. Our final propagators are built from combinations of quark propagators with periodic and anti-periodic boundary conditions in time \cite{Sasaki:2001nf,Detmold:2008yn}. For more details on these so-called ``P+A'' propagators see \cite{Lang:2011mn}.

\subsection{Distillation using Laplacian Heaviside smearing}
  
For an efficient calculation of the quark propagation and flexibility in constructing correlation functions we use the distillation method, first proposed by Peardon {\it et al.} in \cite{Peardon:2009gh}. In this method smeared quark sources and sinks are constructed using a number of low modes of the 3D lattice Laplacian $\nabla^2$. For an $N\times N$ matrix $A$ with eigenvalues $\lambda^{(k)}$ and eigenvectors $v^{(k)}$ one has the spectral decomposition
\begin{equation}
f(A)=\sum_{k=1}^N f(\lambda^{(k)})~v^{(k)}v^{(k)\dagger}.
\end{equation} 
As in \cite{Peardon:2009gh,Lang:2011mn,Lang:2012sv}, the Laplacian-Heaviside (LapH) smearing with $f(\nabla^2)=\Theta(\sigma_s^2+\nabla^2)$ is employed , so the smeared quark fields $q_s$ are 
\begin{align}\label{smearing}
q_s&\equiv \sum_{k=1}^{N} \Theta(\sigma_s^2+\lambda^{(k)}) v^{(k)} v^{(k)\dagger }~q~=~\sum_{k=1}^{N_v} v^{(k)} v^{(k)\dagger }~q~,
\end{align}
where  $N_v$ depends on the target smearing $\sigma_s$. For this study we choose $N_v=96$ or $N_v=64$, depending on the lattice interpolating fields listed in the following sections. The low mode eigenvectors and eigenvalues are calculated using the \texttt{PRIMME} package \cite{Stathopoulos:2009:PPI}.

\subsection{\label{tuning}Tuning the charm quark mass}

For the charm quarks the Fermilab method \cite{ElKhadra:1996mp,Oktay:2008ex} is applied. An approach similar to the method used by the Fermilab Lattice and MILC collaborations \cite{Burch:2009az,Bernard:2010fr} is used and we have previously employed this method to study charmonium and heavy-light mesons in \cite{Mohler:2011ke}. As a slightly modified version is used, the updated procedure is briefly outlined.

In the simplest variant of the Fermilab approach \cite{Burch:2009az,Bernard:2010fr} a single parameter, the charm quark hopping parameter $\kappa_c$ is determined non-perturbatively. To achieve this, the spin-averaged kinetic mass is measured for either charmonium or for heavy-light mesons and its value is tuned to the physical value as determined from experiment. In \cite{Mohler:2011ke} the spin-average of the $1S$ states in the spectrum of $D_s$ mesons was used for this purpose. As we here use gauge configurations with only 2 dynamical quark flavors we instead opt for the $1S$ charmonium states and tune the spin-averaged kinetic mass $(M_{\eta_c}+3M_{J/\Psi})/4$ to its physical value. As further parameters, the Fermilab action contains the clover coefficients $c_E$ and $c_B$ and, incorporating tadpole improvement we choose $c_E=c_B=c_{sw}^{(h)}=\frac{1}{u_0^3}$ where $u_0$ is the average link. To determine the average link we calculate the Landau link on unsmeared gauge configurations. This leads to the numerical value $c_{sw}^{(h)}=1.75218$ for the ensemble used in this study.

With the description outlined above, the remaining task consists of determining the kinetic mass $M_2$ by employing the general form of the lattice dispersion relation from \cite{Bernard:2010fr}
\begin{align}
E(p)&=M_1+\frac{\mathbf{p}^2}{2M_2}-\frac{a^3W_4}{6}\sum_ip_i^4-\frac{(\mathbf{p}^2)^2}{8M_4^3}+ \dots\;,
\label{disp}
\end{align}
where $\mathbf{p}=\frac{2\pi}{L}\mathbf{q}$ for a given spatial extent $L$. Even neglecting higher orders and taking only the terms explicitly listed, this form contains too many parameters to be useful, given the limited number of momentum frames for which a signal of decent statistical quality could be obtained within our setup. We therefore determine $M_2$ using two simplified methods:

\begin{itemize}
\item [(1)] neglect the term with coefficient $W_4$ and fit $M_1$, $M_2$ and $M_4$.
\item [(2)] fit $E^2(p)$ and simplify the $(p^2)^2$ term arising from the mismatch of $M_1$, $M_2$ and $M_4$ by setting $M_1=M_4$ for charmonium and $M_2=M_4$ for heavy-light mesons.
\end{itemize}
Note that (2) differs slightly from the method previously used \cite{Mohler:2011ke}. This change is motivated by the results from method (1) where we obtain $M_1\approx M_4$ for charmonium and $M_2\approx M_4$ for heavy-light mesons. The modified method (2) therefore tests if fits qualitatively change when including the term $-\frac{a^3W_4}{6}\sum_ip_i^4$ breaking rotational symmetry.

For the tuning of $\kappa_c$ we used correlation functions with sources on 2-4 time slices on the full ensemble. A cross check with heavy-light and heavy-strange states used data on 16 time slices. Tables \ref{charm_tuning}, \ref{hs_tuning} and \ref{hl_tuning} list the results for our final choice $\kappa_c=0.123$.

\begin{table}[bht]
\begin{center}
\begin{ruledtabular}
\begin{tabular}{|c|c|c|}
 \T\B & Method (1) & Method (2)\\
\hline
\T\B $M_1$ & 1.52499(42) & 1.52484(42)\\
\hline
\T\B $M_2$ & 1.9581(59) & - \\
\hline
\T\B $M_4$ & 1.5063(216) & - \\
\hline
\T\B $\frac{M_2}{M_1}$ & 1.2840(38) & 1.2745(41)\\
\hline
\hline
\T\B $M_2 [GeV]$ & 3.1186(94)(327) & 3.0951(102)(325)\\
\hline
\hline
\T\B Exp $[GeV]$ & \multicolumn{2}{c|}{$3.06776(30)$}\\
\end{tabular}
\end{ruledtabular}
\end{center}
\caption{\label{charm_tuning}Fit parameters obtained for charmonium with both tuning methods. The values in the last two rows are in GeV, while all other values are in lattice units. The first error on the kinetic mass $M_2$ is statistical while the second error is from the scale setting. The results for $M_4$ are not used in our setup but displayed to demonstrate the observed relation $M_1\approx M_4$. The last row contains the experimental value from \cite{0954-3899-37-7A-075021}.}
\end{table}

\begin{table}[bht]
\begin{center}
\begin{ruledtabular}
\begin{tabular}{|c|c|c|}
 \T\B & Method (1) & Method (2)\\
\hline
\T\B $M_1$ & 1.09704(57) & 1.9716(58)\\
\hline
\T\B $M_2$ & 1.2917(61)& - \\
\hline
\T\B $M_4$ & 1.2397(144) & - \\
\hline
\T\B $\frac{M_2}{M_1}$ & 1.1774(56) & 1.1771(65)\\
\hline
\hline
\T\B $M_2 [GeV]$ & 2.0572(98)(216) & 2.0568(116)(216)\\
\hline
\hline
\T\B Exp $[GeV]$ & \multicolumn{2}{c|}{$2.07634(38)$}\\
\end{tabular}
\end{ruledtabular}
\end{center}
\caption{\label{hs_tuning}Same as Table \ref{charm_tuning} but for charm-strange ($D_s$) mesons.}
\end{table}

\begin{table}[bht]
\begin{center}
\begin{ruledtabular}
\begin{tabular}{|c|c|c|}
 \T\B & Method (1) & Method (2)\\
\hline
\T\B $M_1$ & 1.04226(111) & 1.04206(113)\\
\hline
\T\B $M_2$ & 1.2242(161)& - \\
\hline
\T\B $M_4$ & 1.2550(461) & - \\
\hline
\T\B $\frac{M_2}{M_1}$ & 1.1745(154)& 1.1573(168) \\
\hline
\hline
\T\B $M_2 [GeV]$ & 1.9497(257)(205) & 1.9207(288)(202)\\
\hline
\hline
\T\B Exp $[GeV]$ & \multicolumn{2}{c|}{$1.97140(13)$}\\
\end{tabular}
\end{ruledtabular}
\end{center}
\caption{\label{hl_tuning}Same as Table \ref{charm_tuning} but for charm-light ($D$) mesons. The results for $M_4$ are not used in our setup but displayed to demonstrate the observed relation $M_2\approx M_4$. Notice that the value for $M_2$ in physical units is based on a heavier than physical light-quark mass.}
\end{table}

In all three cases  methods (1) and (2) are in reasonable agreement. As a subset of our charmonium data has been used for tuning $\kappa_c$, it is no surprise that the kinetic mass of the charmonium spin-averaged ground state agrees reasonably well with the PDG value. The spin-averaged $D_s$ mass also compares favorably to the experimental value. While the charmonium result suggests that our charm quark mass has been tuned just a tiny bit to heavy, the result for the heavy-light system comes out somewhat lower than expected for our unphysical light-quark mass. The heavy quark discretization effects for heavy-heavy and heavy-light systems differ and, as we are dealing only with the simplest (lowest order) version of the Fermilab action a perfect agreement is not expected, especially on a rather course lattice. We, however, conclude that our quark masses are tuned reasonably well for the current purpose. This will be tested further in section \ref{charmonium} where the low-lying charmonium spectrum is calculated.

\subsection{Variational method and correlator basis}

To extract the low-lying spectrum we calculate a matrix of correlators at every source  and every sink time slice $t_i$ and $t_f$ 
\begin{align}
C_{ij}(t=t_f-t_i)&=\sum_{t_i}\langle 0|O_i(t_f)O_j^\dagger(t_i)|0\rangle\\
&=\sum_n\mathrm{e}^{-tE_n} \langle 0|O_i|n\rangle \langle n|O_j^\dagger|0\rangle\;,\nonumber
\end{align}
using suitable lattice interpolating fields with definite quantum numbers\footnote{As usual $J$ is the spin, $P$ is parity and $C$ charge conjugation.} $J^{PC}$ (for charmonium) or $J^P$ (for heavy-light states). To extract the low-lying spectrum the generalized eigenvalue problem is solved for each time slice
\begin{align}
C(t)\vec{\psi}^{(k)}&=\lambda^{(k)}(t)C(t_0)\vec{\psi}^{(k)}\;,\\
\lambda^{(k)}(t)&\propto\mathrm{e}^{-tE_k}\left(1+\mathcal{O}\left(\mathrm{e}^{-t\Delta E_k}\right)\right)\;.\nonumber
\end{align}
At large time separation only a single state contributes to each eigenvalue. This procedure is known as the variational method \cite{Luscher:1990ck,Michael:1985ne,Blossier:2009kd}. The employed interpolators are listed in (\ref{MM_0+},\ref{MM_1+}) and the  Appendix \ref{interpolators}.

\section{\label{charmonium}Charmonium results}

Recent lattice simulations of excited charmonium states were presented in \cite{Bali:2011rd,Liu:2012ze,Bali:2011dc,Mohler:2011ke}. The mixing of $c\bar c$ and $D\bar D$ was explored in \cite{Bali:2011rd}, higher spin and exotic $J^{PC}$ states  with carefully determined continuum spin  were presented in  \cite{Liu:2012ze}. 

In our previous study \cite{Mohler:2011ke} the low-lying charmonium states have been determined as a benchmark for our heavy-quark treatment. The resulting low-lying spectrum was in good qualitative agreement with experiment. For our current study, the distillation method based on lowest eigenmodes of the lattice Laplacian enables us to have considerably more freedom with regard to the lattice interpolating fields used. We exploit this opportunity and use an enlarged basis which is tabulated in Appendix \ref{charmonium_appendix}. We consider all non-exotic channels up to spin 3 and also aim at extracting excited energy levels in channels where these are expected above multi-particle thresholds. 

\begin{figure}[tbh]
\includegraphics[width=85mm,clip]{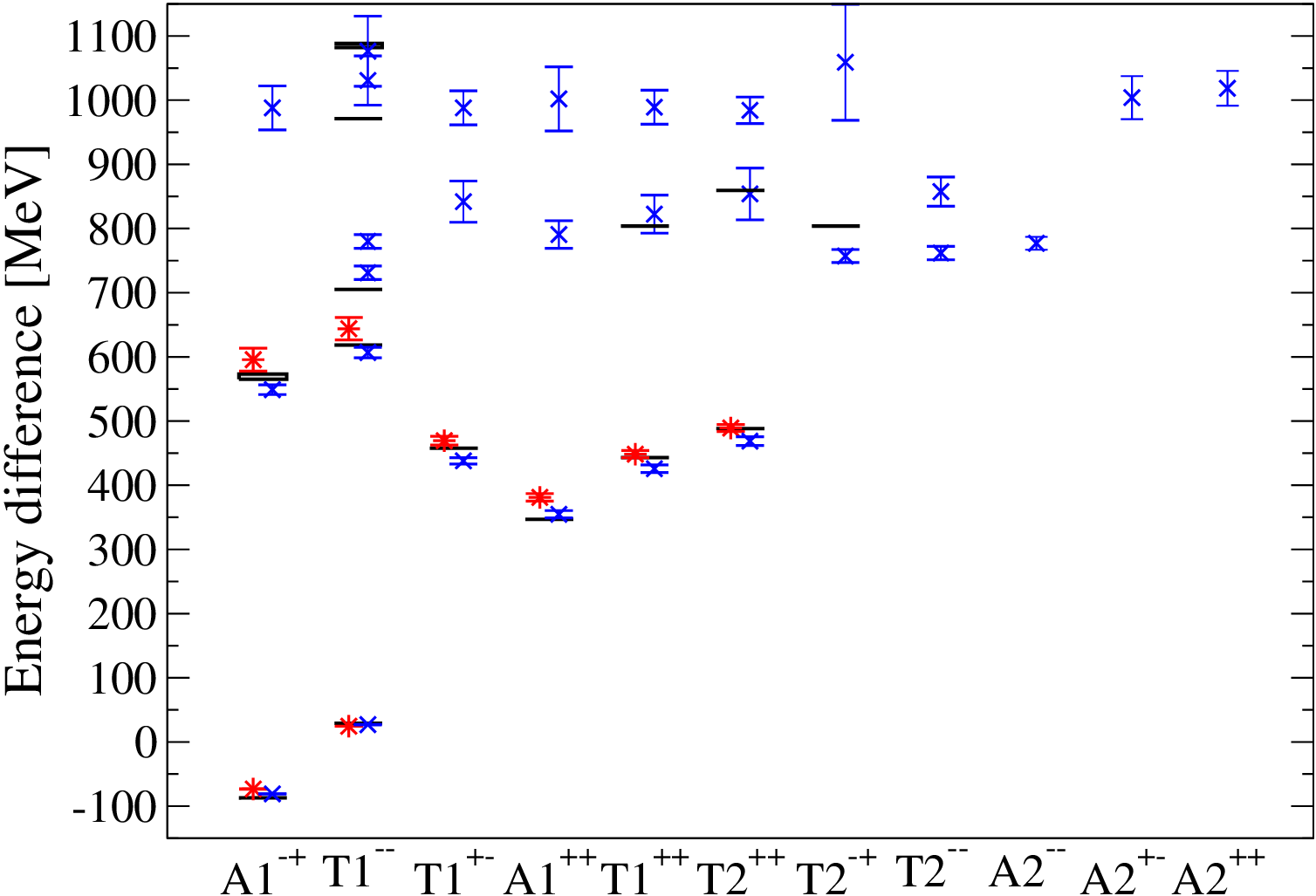}
\caption{ Energy differences $\Delta E=E-\tfrac{1}{4}(M_{\eta_c}+3M_{J\psi})$ for charmonium states on the lattice and in experiment; reference spin-averaged mass is $\tfrac{1}{4}(M_{\eta_c}+3M_{J\psi})\approx 3068~$MeV in experiment.  Levels are listed according to   lattice irreducible representation. 
The results from simulation \cite{Mohler:2011ke} are displayed as red stars and displaced slightly to the left, while our new results displayed as blue crosses are displaced slightly to the right. The statistic and scale setting uncertainties have been combined in quadrature. Experimentally observed states are plotted as black bars or (where there is a substantial uncertainty in mass determination) filled boxes. The level corresponding to the well established $X(3872)$ has been plotted in both the $T1^{++}$ and $T2^{-+}$ irreps, reflecting its uncertain quantum numbers \cite{delAmoSanchez:2010jr,Choi:2011fc}.} 
\label{charm_results_irrep}
\end{figure}

Figure \ref{charm_results_irrep} collects results for all lattice irreps and non-exotic quantum numbers. For all states the difference with respect to the spin-averaged ground state $M_{\overline{1S}}=(M_{\eta_c}+3M_{J/\Psi})/4$ is plotted. The use of a larger basis enables us to extract at least the ground state in all channels investigated and in many cases also one or more excited state energy levels can be extracted. Details of the fitting methodology can be found in the appendix. In addition to our new data (blue crosses) we also display the results from \cite{Mohler:2011ke} (red stars).

\begin{figure}[tbh]
\includegraphics[width=85mm,clip]{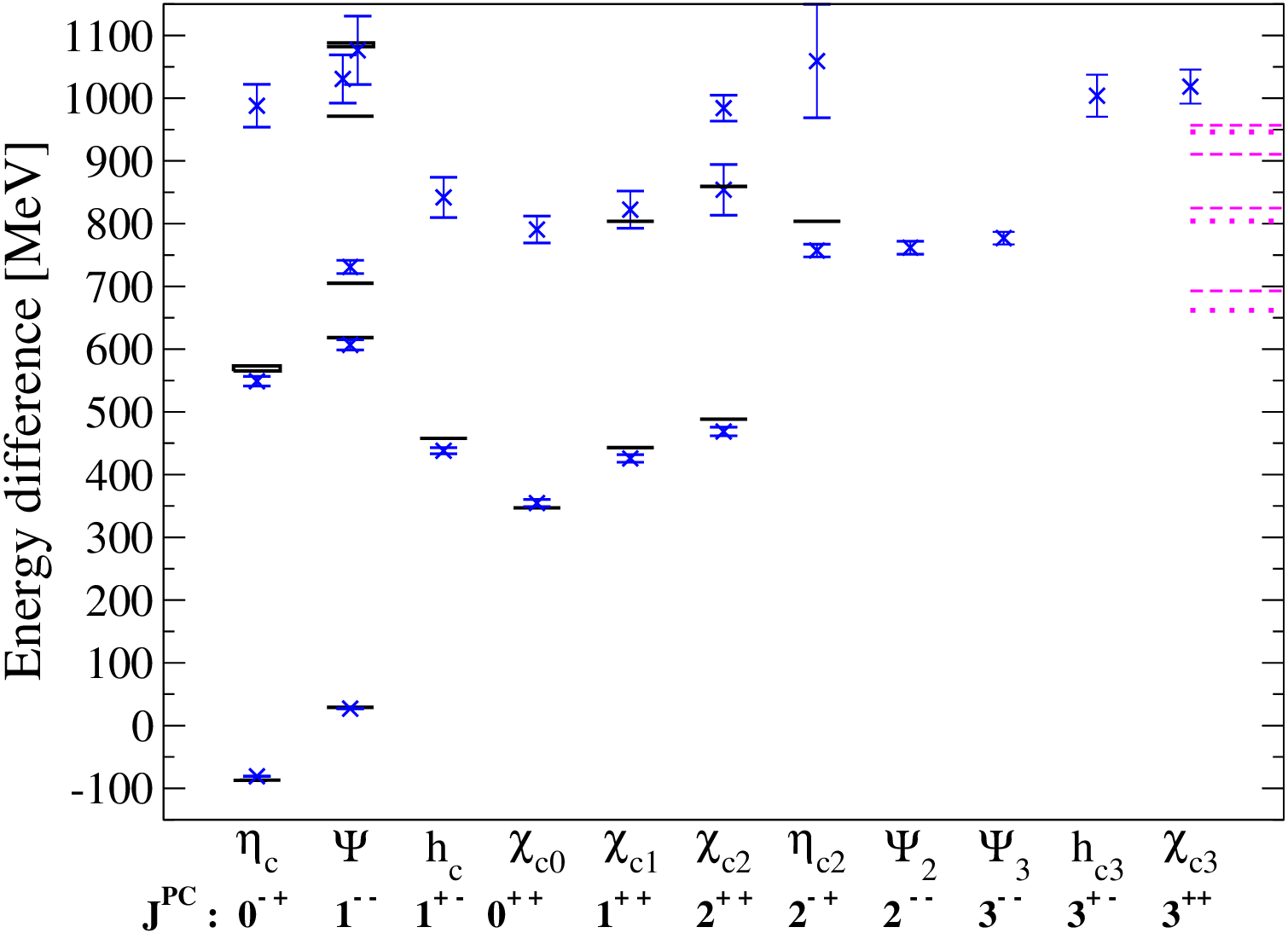}
\caption{Energy differences $\Delta E=E-\tfrac{1}{4}(M_{\eta_c}+3M_{J\psi})$ for charmonium states on our lattice and in experiment; reference spin-averaged mass is $\tfrac{1}{4}(M_{\eta_c}+3M_{J\psi})\approx 3068~$MeV in experiment.
In addition to the low-lying 1S, 1P and 2S states, the full set of 1D and 2P states is also seen. The black lines or boxes denote known experimental states with uncertainty on their masses. The level corresponding to  $X(3872)$ is plotted for both possible quantum numbers $1^{++}$ and $2^{-+}$.  The magenta lines on the right denote relevant lattice and continuum thresholds. From low to high energy, the  physical $DD$, $D^* D$ and $D^* D^*$ thresholds are plotted as dotted lines. The corresponding lattice thresholds are plotted as dashed lines in a similar order: $DD$ in S-wave, $D^* D$ in S-wave, $DD$ in P-wave and $D^* D^*$ in S-wave. The corresponding $D_s$ meson thresholds are omitted as calculations were performed with a 2-flavor sea.  } 
\label{charm_results_channel}
\end{figure}

Figure \ref{charm_results_channel} shows the same energy levels as before, assigned to continuum states. For the assignment, degeneracies across irreps and interpolator overlaps have been used\footnote{For a more elaborate way to identify the continuum spin please refer to \cite{Dudek:2009qf,Dudek:2010wm}}, including data from the E irreducible representation which is not shown in Figure \ref{charm_results_irrep}. 

The dotted lines in Figure \ref{charm_results_channel} denote the physical $DD$, $D^* D$ and $D^* D^*$ thresholds. Our results for states below all thresholds agree well with the experimental states, which are commonly interpreted as the $1S$, $1P$ and $2S$ multiplets (from low to high mass). Above the physical $DD$ threshold we observe another band of states of quantum numbers $1^{--}$, $2^{--}$, $3^{--}$ and $2^{-+}$ which is naturally interpreted as the $1D$ states. Notice that we seem to observe the D-wave $3^{--}$ state also in the $T1^{--}$ irrep. The $J^{PC}=1^{--}$ D-wave state corresponds to the experimental $\Psi(3770)$ which has an appreciable decay into two D mesons. Note that this decay proceeds in P-wave and the corresponding lattice threshold is far away from the physical $DD$ threshold. As a result we do not expect to reproduce the correct $\Psi(3770)$ mass or the correct splitting between the $\Psi(2S)$ and $\Psi(3770)$. Similar remarks can be made for all other states above threshold, so we restrict ourselves to some qualitative observations. Of particular interest is the observation of a further band of states split from the spin averaged $1S$ states by around 850 MeV in our simulation. These states have the pattern expected for the $2P$ states. Notice that the $\chi_{c2}(2P)$ has been identified in experiment and the corresponding energy level is included in our plot. Furthermore, a recent study by BaBar \cite{Lees:2012xs} suggests that the $X(3915)$ is likely to have quantum numbers $0^{++}$ and could be interpreted as the $\chi_{c0}(2P)$. The identification of this state with the $\chi_{c0}(2P)$ is not appealing on theoretical grounds \cite{Guo:2012tv} and there are indications for a broad $\chi_{c0}(2P)$ \cite{Guo:2012tv} at a lower mass, which would be more compatible with our data\footnote{For the splitting between the $\chi_{c2}(2P)$ and the $\chi_{c0}(2P)$ we obtain $63\pm33$ MeV.}. Evidence for the $\Psi_2$ with quantum numbers $J^P=2^{--}$ has furthermore been found by the Belle collaboration \cite{talk:belle} at a mass of $3823.5(2.8)$ MeV. With regard to the  X(3872), whose quantum numbers are not yet settled \cite{delAmoSanchez:2010jr,Choi:2011fc} and could be either $1^{++}$ or $2^{-+}$, we find that one of the states we interpret as a 2P state is close to the mass of the X(3872). Just like in experiment, this state is compatible in mass with the $D^* D$ threshold. It should be stressed that the current results are within uncertainties compatible with both possible quantum numbers and that one can not draw any strong conclusion about the nature of the X(3872) from this study. Nevertheless it is interesting that a state is observed in close vicinity to the threshold in the $1^{++}$ channel, while the ground state in the $2^{-+}$ channel comes out lighter. For a recent lattice study investigating this issue see \cite{Yang:2012my}. For a particularly insightful discussion regarding the possible nature of the X(3872) see \cite{Artoisenet:2010va}.

In addition we observe a number of further states for which likely assignments are shown in the figure. In particular we find two spin 3 F-wave states and another set of excited S-waves.

To disentangle spin-dependent from spin-independent contributions we further define spin-averaged masses
\begin{align}
M_{\overline{2S}}&=\frac{1}{4}(M_{\eta_c^\prime}+3M_{J/\Psi^\prime})\;,\nonumber\\
M_{\overline{1P}}&=\frac{1}{9}(M_{\chi_{c0}}+3M_{\chi_{c1}}+5M_{\chi_{c2}})\;,\\
M_{\overline{2P}}&=\frac{1}{9}(M_{\chi_{c0}^\prime}+3M_{\chi_{c1}^\prime}+5M_{\chi_{c2}^\prime})\nonumber\;.
\end{align}
The results are listed in Table \ref{charmonium_table}. In addition we take a look at the hyperfine splittings between spin-singlet and spin-triplet states
\begin{align}
M_{n^3L}-M_{n^1L}
\end{align}
and at the P-wave spin-orbit and tensor splittings
\begin{align}
M_{Spin-Orbit}&=\frac{1}{9}(5M_{\chi_{c2}}-3M_{\chi_{c1}}-2M_{\chi_{c0}})\;,\\
M_{Tensor}&=\frac{1}{9}(3M_{\chi_{c1}}-M_{\chi_{c2}}-2M_{\chi_{c0}})\;.\nonumber
\end{align}
Depending on the heavy quark treatment, lattice discretization effects in these quantities can be substantial. Their determination is a challenge for lattice QCD. Values extracted for the $1S$, $2S$ and $1P$ hyperfine splittings and for the $1P$ and $2P$ spin-orbit and tensor splittings are presented in Table \ref{charmonium_table}. The experiment values in the table are the corresponding PDG values \cite{0954-3899-37-7A-075021}. In the case of the $1S$ $\eta_c$ state which enters the hyperfine splitting, the PDG average has a poor confidence level and newer results from BESIII \cite{BESIII:2011ab} and Belle \cite{Zhang:2012tj} suggest that the hyperfine splitting is substantially lower. For recent lattice results on the $1S$ hyperfine splitting including a continuum extrapolation see \cite{Becirevic:2012dc,Briceno:2012wt,Donald:2012ga}.

For the charmonium hyperfine splitting we also determine the uncertainty associated with the kappa tuning procedure outlined in Section \ref{tuning}.  First we average over the results from method (1) and method (2) for our tuning runs at $\kappa_c=0.123$ and $\kappa_c=0.124$ which are close to the value corresponding to physical $M_2$. We determine the resulting values for $M_2$ and the hyperfine splitting for both values of $\kappa_c$. Due to the enlarged statistics our final charmonium data differs slightly from the tuning run at the same $\kappa_c$ and we also determine the kinetic mass $M_2$ for our final data. We then use interpolations of the tuning data to determine the kappa tuning uncertainty for our final data.  For this purpose the uncertainty from our choice of fitting model is estimated by the difference between the kinetic masses obtained from method (1) and method (2). For the total error, the stochastic error from the Monte Carlo estimation, the scale setting uncertainty and the uncertainty from the fitting model are added in quadrature. As our final run turns out to have slightly missed the physical $M_2$ we obtain an asymmetric error of $\pm_{0.0}^{2.2}$ from the uncertainty in the charm quark mass. Similar tuning errors of about 2\% are expected for all spin-dependent quantities.

In section \ref{tuning} we also took a look at the kinetic masses of $D$ and $D_s$ mesons. In addition to charmonium mass splittings the values for the combinations $2M_{\overline{D}}-M_{\overline{c\bar{c}}}$ and $2M_{\overline{D_s}}-M_{\overline{c\bar{c}}}$ are also provided in Table \ref{charmonium_table}. In these combinations the leading heavy quark contribution drops out. Again the proximity of our results to the experimental values is encouraging.

\begin{table}[bht]
\begin{center}
\begin{ruledtabular}
\begin{tabular}{|c|c|c|}
 \T\B Mass difference & This paper [MeV] & Experiment [MeV]\\
\hline
\hline
$\chi_{c0}-\overline{1S}$ & $354.7\pm4.4\pm3.7$ & $347.0\pm0.4$\\
\hline
$\chi_{c1}-\overline{1S}$ & $425.7\pm3.9\pm4.7$ & $442.9\pm0.3$\\
\hline
$\chi_{c2}-\overline{1S}$ & $468.7\pm4.7\pm4.9$ & $488.4\pm0.3$ \\
\hline
$h_c-\overline{1S}$ & $438.0\pm4.9\pm4.6$ & $457.7\pm0.4$ \\
\hline
$\overline{1P}-\overline{1S}$ & $441.7\pm4.0\pm4.6$ & $457.5\pm0.3$\\
\hline
$\eta_c^\prime-\overline{1S}$ & $548.9\pm4.9\pm5.8$ &  $569.2\pm4.0$\\
\hline
$J/\Psi^\prime-\overline{1S}$ & $606.8\pm4.9\pm6.4$ &  $618.3\pm0.3$\\
\hline
$\overline{2S}-\overline{1S}$ & $592.3\pm4.9\pm6.2$ &  $606.1\pm1.0$\\
\hline
1S hyperfine & $107.9\pm0.3\pm1.1\pm_{0.0}^{2.2}$ & $116.6\pm1.2$\\
\hline
1P spin-orbit & $39.7\pm2.1\pm0.4$ &  $46.6\pm0.1$ \\
\hline
1P tensor & $11.02\pm0.87\pm0.12$ &  $16.25\pm0.07$ \\
\hline
1P hyperfine & $3.7\pm2.7$& $-0.10\pm0.22$ \\
\hline
2S hyperfine & $57.9\pm2.0$ & $49\pm4$ \\
\hline
2P spin-orbit & $24.6\pm15.7\pm0.3$ & - \\
\hline
2P tensor & $2.2\pm4.3$ & - \\
\hline
$\overline{2P}-\overline{1S}$ & $836.4\pm30.5\pm8.8$ & - \\
\hline
$2M_{\overline{D}}-M_{\overline{c\bar{c}}}$ & $890.9\pm3.3\pm9.3$ & $882.4\pm0.3$ \\
\hline
$2M_{\overline{D_s}}-M_{\overline{c\bar{c}}}$ & $1065.5\pm1.4\pm11.2$ & $1084.8\pm0.6$ \\
\end{tabular}
\end{ruledtabular}
\end{center}
\caption{\label{charmonium_table} Mass differences in the charmonium spectrum in MeV compared to experimental values (calculated from \cite{0954-3899-37-7A-075021}; the value for the 1P hyperfine splitting is from \cite{Olsen:2012xn}). Bars denote spin-averaged values. For the results of this paper, the first error denotes the statistical uncertainty and the second error denotes the uncertainty from setting the lattice scale\footnote{For spin-dependent quantities the indirect contribution of the scale setting uncertainty to the kappa tuning uncertainty is sizable. Our scale setting error only accounts for the direct uncertainty associated with the setting of the lattice scale.}. 
In addition there is a non-negligible error from the uncertainty in the determination of $\kappa_c$ for all spin-dependent quantities. For the 1S hyperfine-splitting the corresponding error is estimated and given as the third (asymmetric) error. It is stressed that the gauge ensembles at our disposal do not allow for a continuum and infinite volume extrapolation. Consequently qualitative but not quantitative agreement is expected. In the last line we also provide the splitting $2M_{\overline{D_s}}-M_{\overline{c\bar{c}}}$ which can be directly compared to the results quoted by the Fermilab lattice and MILC collaborations \cite{Burch:2009az} and also to the value of $2M_{D_s}-M_{\eta_c}$ quoted by HPQCD in \cite{Follana:2006rc}. For the determination of the strange quark mass on our lattices please refer to \cite{Lang:2012sv}.}
\end{table}

\section{\label{dmesons}D mesons resonances including meson-meson interpolating fields}

\subsection{Energy levels and the L\"uscher method}

This section provides a short but general introduction for the extraction of resonance parameters. A small modification specific to our heavy-quark setup is discussed in the following section.

Assuming a localized interaction, the energy levels of a two-hadron system in a finite box are related to the scattering phase shift in the elastic region \cite{Luscher:1985dn,Luscher:1986pf,Luscher:1990ux,Luscher:1991cf}. One first needs to determine the energy levels $E$ of the two-hadron system on the lattice. We choose the total momentum of the system to be zero in this simulation, so the lattice frame coincides with in the center of momentum (CM) frame, and both hadrons have momentum $p^*$. In this case we avoid the complications that arise for the extraction of the S-wave from simulations with non-zero total momentum caused by $m_D^{(*)}\ne m_\pi$ \cite{Fu:2011xz,Leskovec:2012gb}. In the exterior region, where the interaction is negligible,  
\begin{align}\label{reldisp}
E^2&=s\\
&=\left(\sqrt{{p^*}^2+m_{H_1}^2}+\sqrt{{p^*}^2+m_{H_2}^2}\right)^2\;,\nonumber
\end{align}
and the  discrete values of $p^*$ are extracted from $E$ via
\begin{align}
{p^*}^2&=\frac{1}{4s}\left(s-(m_{H_1}+m_{H_2})^2\right)\left(s-(m_{H_1}-m_{H_2})^2\right)\;,
\label{pstardef}
\end{align}
while the corresponding dimensionless momentum $q$ is defined as
\begin{align}
q=\frac{L}{2\pi}~p^*\;.
\end{align}

Determining the value of the momentum $q$ from these relations, one obtains the relevant S-wave scattering phase shift  $\delta_0$ from the 
L\"uscher formula \cite{Luscher:1990ux}
\begin{align}
\label{luscher_eq}
\tan{\delta_0(q)}&=\frac{\pi^{\frac{3}{2}}q}{\mathcal{Z}_{00}(1;q^2)}\;,
\end{align}
which applies for total momentum zero. Here $\mathcal{Z}_{00}(1;q^2)$ is a generalized zeta function. This relation neglects higher partial waves $\delta_{l\geq 4}$ in the case of $D\pi$ scattering with $J^P=0^+$ ($A_1^+$ irrep of $O_h$), and it neglects  $\delta_{l=2}$ in the case of $D^*\pi$ scattering with $J^P=1^+$ ($T_1^+$ irrep). It also neglects terms exponentially suppressed with the lattice volume, and we note that these terms might not be completely negligible for our volume. We are setting up for simulations at a larger volume and results will indicate whether this might affect our present results.

The elastic scattering phase $\delta$ is related to the scattering amplitude $T_l$ by
\begin{align}
T_l=\sin \delta_l~\mathrm{e}^{i\delta_l}=\frac{\mathrm{e}^{2i\delta_l}-1}{2i}\;.
\end{align}
A variable $\rho_l(s)$, defined as
\begin{equation}
\label{rho}
\rho_l(s) \equiv \frac{(p^*)^{2l+1}\cot\delta(p^*)}{\sqrt{s}}\;,
\end{equation}
depends on the scattering length $a_l$ near threshold
\begin{align}
\label{a0}
\sqrt{s}\rho_l(s)&=\frac{1}{a_l}+\mathcal{O}({p^*}^2)\;.
\end{align}

For the case of an elastic channel dominated by a single resonance one can also assume a relativistic Breit-Wigner amplitude and obtain
\begin{align}
\label{bwtype}
&\sqrt{s}\Gamma_r(s)~\cot\delta(q)=s_r-s\;,\\
&\Gamma_r(s)=g^2\frac{(p^*)^{2l+1}}{s}\qquad s_r=m_r^2\;,\nonumber
\end{align} 
where the width of the resonance $\Gamma_r$ has been parametrized by a phase space factor and a coupling constant $g$; in our case $g$ will either be $g_{D_0^* D\pi}$ or $g_{D_1D^*\pi}$. The momentum dependence of $\Gamma_r$ ensures that the amplitude has the correct threshold behavior in the elastic region and defines how the amplitude is continued below threshold. The phase shift vanishes as $p^*$ goes to zero and below threshold the amplitude is real.
We will extract the mass $m_r$ and the coupling $g$ from the resulting $\rho_l(s)$ via
\begin{equation}
\label{BW_rho}
\rho_l(s)=\frac{1}{g^2}(s_r-s)
\end{equation}
 obtained by combining Eqs. (\ref{bwtype}) and (\ref{rho}). 

\subsection{Dispersion relations}

\begin{figure}[tb]
\includegraphics[width=80mm,clip]{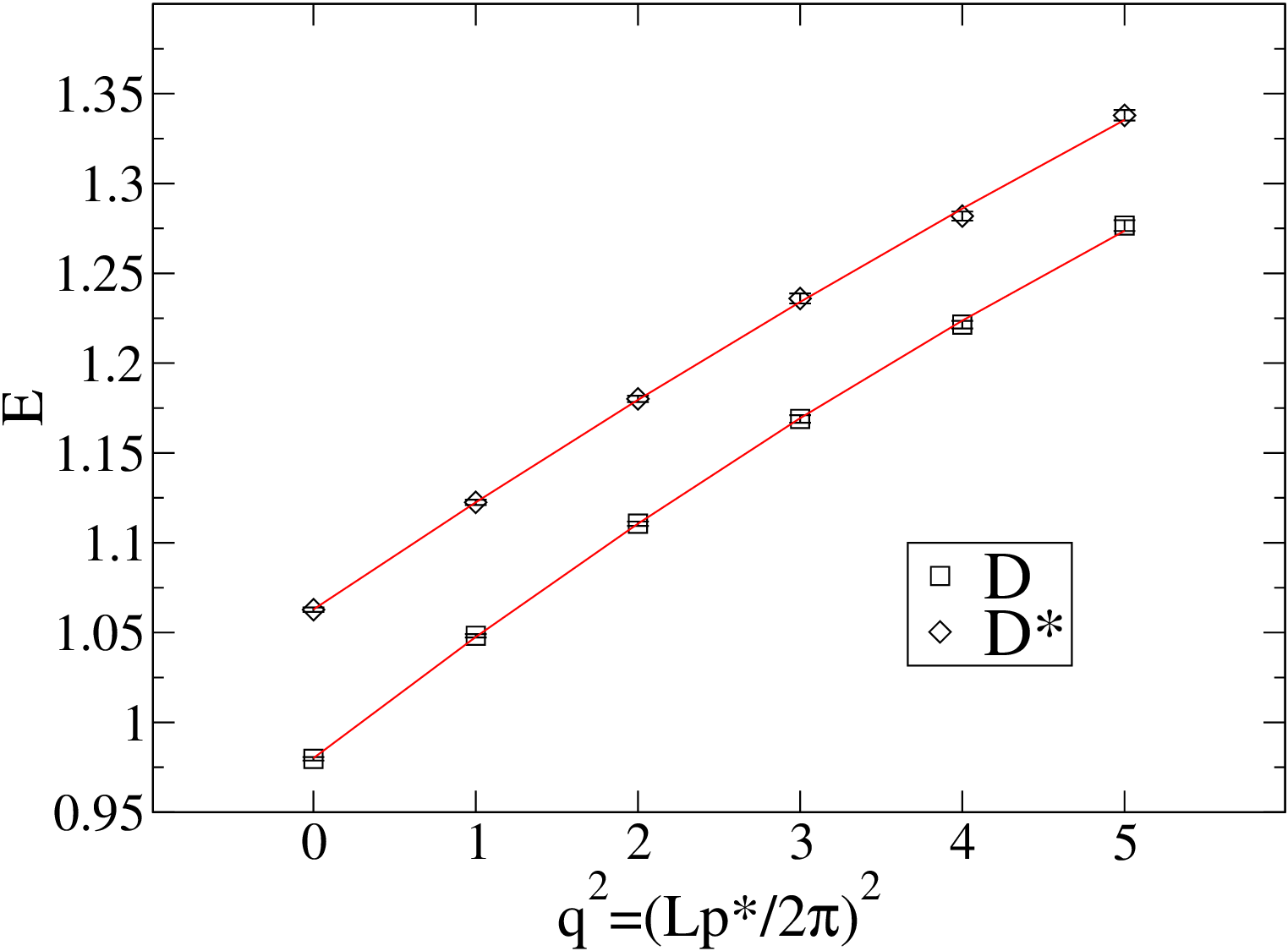}
\caption{Measured energies of $D$ or $D^*$ as a function of $q^2=(L p^*/2\pi)^2$ and a fit using the dispersion relation (\ref{disp}) with method (1):  $M_1\approx 0.980$, $M_2\approx 1.107$, $M_4\approx 1.107$ are extracted for $D$, while $M_1\approx1.063$, $M_2\approx 1.267$, $M_4\approx 1.325$ are extracted for $D^*$.} 
\label{dispersion_D_Dstar}
\end{figure}

As already mentioned we have to modify the above procedure slightly to account for our unphysical dispersion relation. Equation (\ref{reldisp}) uses a relativistic dispersion relation for both hadrons. In our case one hadron is a pion, where the relativistic continuum dispersion relation holds well \cite{Lang:2012sv} in the simulation. The other hadron is a heavy-light $D$ or $D^*$ meson with the  dispersion relation (\ref{disp}) displayed in Fig. \ref{dispersion_D_Dstar}, where the values for $M_1$, $M_2$ and $M_4$ are determined separately for $D$ and $D^*$ with method (1) and provided  in the caption. Therefore, we extract the  momentum $p^*$ from the energy $E$ of the $D^{(*)}\pi$ system  via
\begin{equation}
E=\sqrt{m_\pi^2+{p^*}^2}+ M_1+\frac{{p^*}^2}{2M_2}-\frac{{p^*}^4}{8M_4^3}\:.
\end{equation}

For  convenience the values of separate energies for $\pi$, $D$ and $D^*$ at momenta $q=0$ and $q=1$ are listed in Table \ref{gstable}. Their sums provide the reference energies of the lowest two non-interacting scattering states and are plotted as dashed lines in Figs. \ref{d0meson_effmass} and \ref{d1meson_effmass}.   

\begin{table}[thb]
\begin{ruledtabular}
\begin{tabular}{ccc}
\T\B meson & $Ea~(p^*=0)$ & $E^{d.r.}a~(p^* =\tfrac{2\pi}{L})$\\
\hline
\T\B $\pi$ & 0.1673(16) & 0.4268(65) \\
\T\B $D$ & 0.9801(10) & 1.0476(10)\\
\T\B $D^*$ & 1.0629(13) & 1.1225(14) \\
\hline
\end{tabular}
\end{ruledtabular}
\caption{\label{gstable}Energies of $\pi$, $D$ and $D^*$ for momenta $p^*=0~,\tfrac{2\pi}{L}$, which are relevant for non-interacting $D^{(*)}(0)\pi(0)$ and  $D^{(*)}(1)\pi(-1)$. The pion energy at $p^*=\tfrac{2\pi}{L}$ is based on the continuum dispersion relation; the energies of $D$ and $D^*$ are obtained with the dispersion relation (\ref{disp}) within method (1) and with $M_{1,2,4}$ from Fig. \ref{dispersion_D_Dstar}.} 
\end{table}

\subsection{Results}

For the $D$ mesons a basis consisting of quark-antiquark and meson-meson interpolating fields is used. For the $q\bar{q}$ part the interpolating fields are tabulated in Table \ref{interpolators_dmesons}. For mesons made from  quarks with different  masses, charge conjugation (or more generally G-parity) is not a good quantum number. In case of the $J^P=1^+$ $D_1$ meson we therefore consider also mixing between interpolating fields corresponding to different charge conjugations in the mass-degenerate case. This has already been found to be important in \cite{Mohler:2011ke} and has been investigated for kaons in \cite{Engel:2011aa}. 

For the $I=\frac{1}{2}$ resonances in the $J^P=0^+$ ($D_0^*$) and $J^P=1^+$ ($D_1$) channels, one needs the following meson-meson combinations
\begin{align}
\left|\frac{1}{2},\frac{1}{2}\right>_{\bar D_0^{* 0}}&=\sqrt{\frac{2}{3}}D^-\pi^++\sqrt{\frac{1}{3}}\bar D^0\pi^{0}\;,\\
\left|\frac{1}{2},\frac{1}{2}\right>_{\bar D_1^0}&=\sqrt{\frac{2}{3}}D^{* -}\pi^++\sqrt{\frac{1}{3}}\bar D^{* 0}\pi^{0}\;,\nonumber
\end{align}
with
\begin{align}
\bar D^0&=\bar{c}\Gamma u\;,\nonumber\\
D^-&=\bar{c}\Gamma d\;,\\
\pi^+&=\bar{d}\Gamma u\;,\nonumber\\
\pi^0&=\frac{1}{\sqrt{2}}\left(\bar{u}\Gamma u-\bar{d}\Gamma d\right)\;,\nonumber
\end{align}
and $\Gamma=\gamma_5,~ \gamma_i$ for pseudoscalar and vector fields, respectively. 
In case of the $D_0^*$ we use a basis of six interpolating fields in irrep $A_1^+$ of $O_h$, which couples to $J^P=0^+$ (with negligible contributions from $J,l\geq 4$ due to broken rotational symmetry). The first four are $q\bar{q}$ interpolators as listed in Table \ref{interpolators_dmesons} and the last two are meson-meson interpolators:
\begin{align}
\label{MM_0+}
\mathcal{O}_5(t)&=\sqrt{\frac{2}{3}}D^-(0)\pi^+(0)+\sqrt{\frac{1}{3}}\bar D^0(0)\pi^{0}(0)\;,\\
\mathcal{O}_6(t)&=\sum_i\left[\sqrt{\frac{2}{3}}D^-(\mathbf{p}_i)\pi^+(-\mathbf{p}_i)+\sqrt{\frac{1}{3}}\bar D^0(\mathbf{p}_i)\pi^{0}(-\mathbf{p}_i)\right]\;.\nonumber
\end{align}
Interpolator $\mathcal{O}_6$ is constructed from nontrivial momenta $\frac{2\pi}{L}\mathbf{p_i}$ with
\begin{align}
\mathbf{p}_1&=(1,0,0)\;,\qquad&\mathbf{p}_4=(-1,0,0)\;,\nonumber\\
\mathbf{p}_2&=(0,1,0)\;,\qquad&\mathbf{p}_5=(0,-1,0)\;,\\
\mathbf{p}_3&=(0,0,1)\;,\qquad&\mathbf{p}_6=(0,0,-1)\;.\nonumber
\end{align}
For the $D_1$ we use a basis of ten interpolating fields in irrep $T_1^+$ of $O_h$, which couples to $J^P=1^+$ (and $J\geq 3$).  Again just two of these are meson-meson interpolators constructed in a similar way
\begin{align}
\label{MM_1+}
\mathcal{O}_9(t)&=\sqrt{\frac{2}{3}}D_k^{* -}(0)\pi^+(0)+\sqrt{\frac{1}{3}}\bar D_k^{* 0}(0)\pi^{0}(0)\;,\\
\mathcal{O}_{10}(t)&=\sum_i\left[\sqrt{\frac{2}{3}}D_k^{* -}(\mathbf{p}_i)\pi^+(-\mathbf{p}_i)+\sqrt{\frac{1}{3}}\bar D_k^{*0}(\mathbf{p}_i)\pi^{0}(-\mathbf{p}_i)\right]\;,\nonumber
\end{align}
where the $D^*_k=\bar c\gamma_k q$ polarization is along $k$ and correlators are averaged over $k=x,y,z$ in the end. In the case of the interpolating fields with non-trivial momenta, we restrict the number of Laplacian eigenmodes used in the construction to $N_v=64$, while $N_v=96$ is used for all other interpolating fields. The contractions for such $I=1/2$ interpolators are explicitly provided in Appendix B of the $K\pi$ scattering simulation \cite{Lang:2012sv}, with the only necessary replacement $\bar{s}\to \bar{c}$. 

The fitting methodology is the same as for charmonium and is outlined in Appendix \ref{fit_met}. Table \ref{fit_details_dmesons} in Appendix \ref{dmesons_appendix} lists our choices for the interpolator sets, timeslice $t_0$ and fit ranges as well as the fit results and $\chi^2/$d.o.f. for the fit in the channels without meson-meson interpolating fields; the basis used is indicated by the interpolator numbers from Table \ref{interpolators_dmesons}. The energy levels for these channels are collected in Figures \ref{dmeson_results_channel} and \ref{dmeson_results_final}. The results for the $D_0^*$ and $D_1$ channels that take into account the meson-meson  interpolators are discussed separately below.

\subsubsection{$D\pi$ scattering in the $J^P=0^+$ channel and $D_0^*$ resonance}

\begin{table*}[t]
\begin{ruledtabular}
\begin{tabular}{cccccccccc}
\T\B level $n$ & interpolators & $t_0$ & fit range  & type & $E_na$ & $\chi^2/$d.o.f & $ap^*$ & $sa^2$ & $\delta_0$\\ 
\hline
1 & 1,3,5,6 & 3 & 4-21 & 2 exp & 1.1145(25) & 3.39/13 & 0.0939(34)i & 1.2420(65)  & 41.2(12.2)i\\
2 & 1,3,5,6 & 3 & 4-13 & 2 exp & 1.3060(52) & 4.73/6 & 0.2474(51) & 1.7057(135) & -77.1(2.8)\\
3 & 1,3,5,6 & 3 & 4-11 & 2 exp & 1.495(15)  & 0.35/4 & 0.4093(118) & 2.236(45)   & -16.7(12.4)\\
\end{tabular}
\end{ruledtabular}
\caption{\label{d0results}Final results for the lowest three energy levels in the  $D_0^*$ channel with $J^P=0^+$. For each state the timeslice $t_0$ for the variational method, the fit range, fit type and $\chi^2$/d.o.f as well as results for the energy $E_n$, the momentum $p^*$ defined in (\ref{pstardef}), the invariant mass squared $s$ and the S-wave scattering phase $\delta_0$ are provided. Interpolators ${\cal O}_{1-4}$ of type $u\bar c$ are listed in Table \ref{interpolators_dmesons}, while ${\cal O}_{5,6}$ of type $D\pi$ are given in (\ref{MM_0+}).}
\end{table*}

\begin{table*}[t]
\begin{ruledtabular}
\begin{tabular}{cccccccccc}
\T\B level $n$ & interpolators & $t_0$ & fit range & type & $E_na$ & $\chi^2/$d.o.f & $ap^*$ & $sa^2$ & $\delta_0$\\ 
\hline
1 & 1,4,7,8,9,10 & 3 & 9-19 & 1 exp & 1.1978(28) & 7.30/9 & 0.0938(40)i  & 1.4348(68) & 40.9(14.0)i\\
2 & 1,4,7,8,9,10 & 3 & 8-16 & 1 exp & 1.3222(90) & 1.72/7 & 0.1810(291) & 1.748(24)  &     /        \\
3 & 1,4,7,8,9,10 & 3 & 8-16 & 1 exp & 1.3456(71) & 5.34/7 & 0.2068(80)  & 1.811(19)  & -55.7(4.0)\\
4 & 1,4,7,8,9,10 & 3 & 7-11 & 1 exp & 1.571(10)  & 0.30/3 & 0.4107(83)  & 2.469(32)  & -16.1(7.0)\\
\end{tabular}
\end{ruledtabular}
\caption{\label{d1results}Final results for the lowest four energy levels in the $D_1$ channel with $J^P=1^+$. For an explanation of the entries see Table \ref{d0results}. S-wave phase shifts $\delta_0$ are extracted under assumptions given in the main text: levels $1,3,4$  are assumed to be dominated by $\delta_0$, while level $2$  corresponds to a narrow  $D^*\pi$ resonance in D-wave ($\delta_0$ is therefore not provided for this level).}
\end{table*}

\begin{figure}[tb]
\includegraphics[width=85mm,clip]{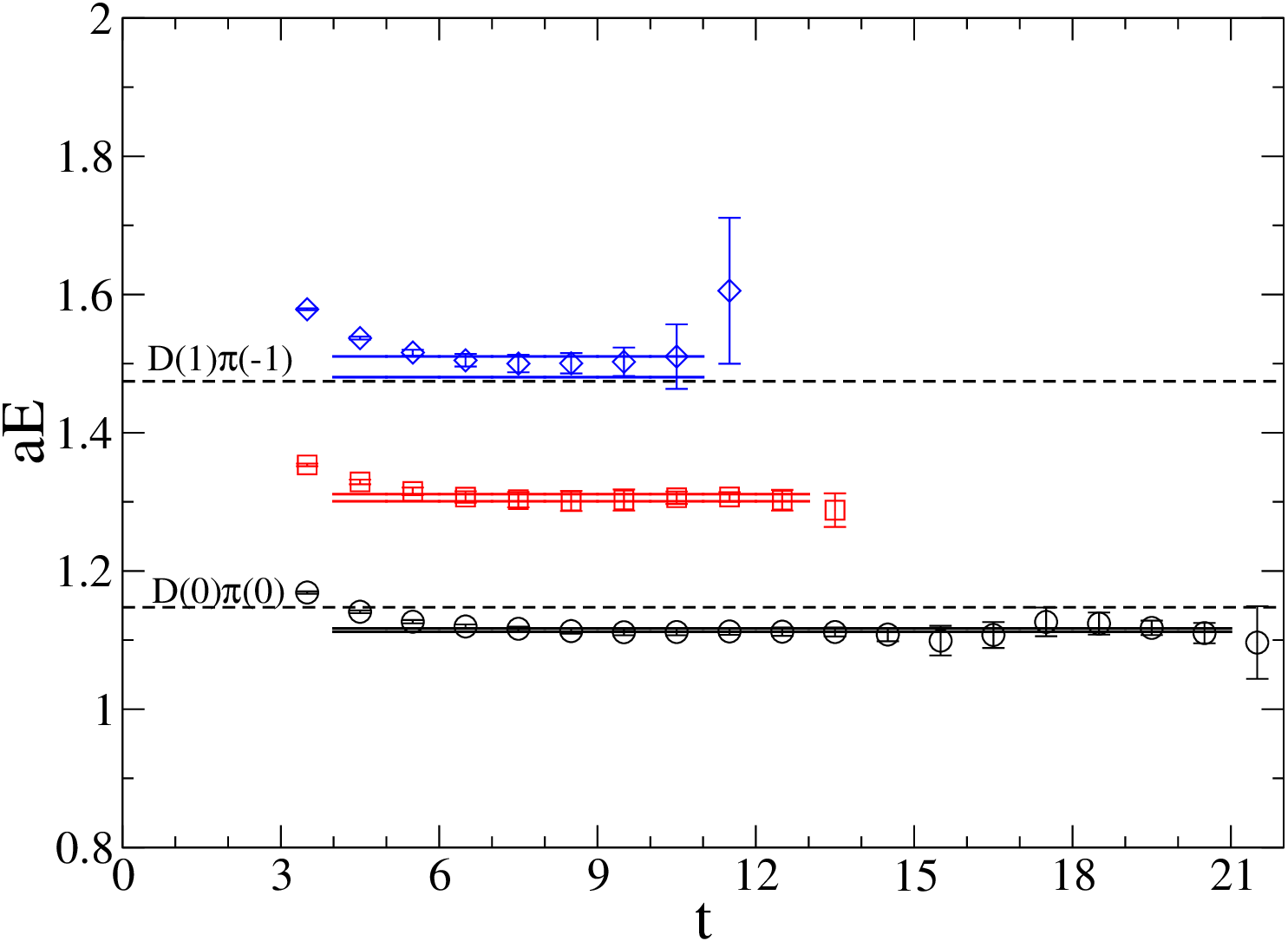}
\caption{Effective masses of the lowest three energy levels in the $D_0^*$ channel with $J^P=0^+$. The fit ranges and uncertainties are indicated by the solid horizontal lines. In addition, non-interacting scattering levels $D(p)\pi(-p)$ are depicted by dashed lines. Interpolator choices and numerical values can be found in Table \ref{d0results}.} 
\label{d0meson_effmass}
\end{figure}

\begin{figure}[tb]
\includegraphics[width=85mm,clip]{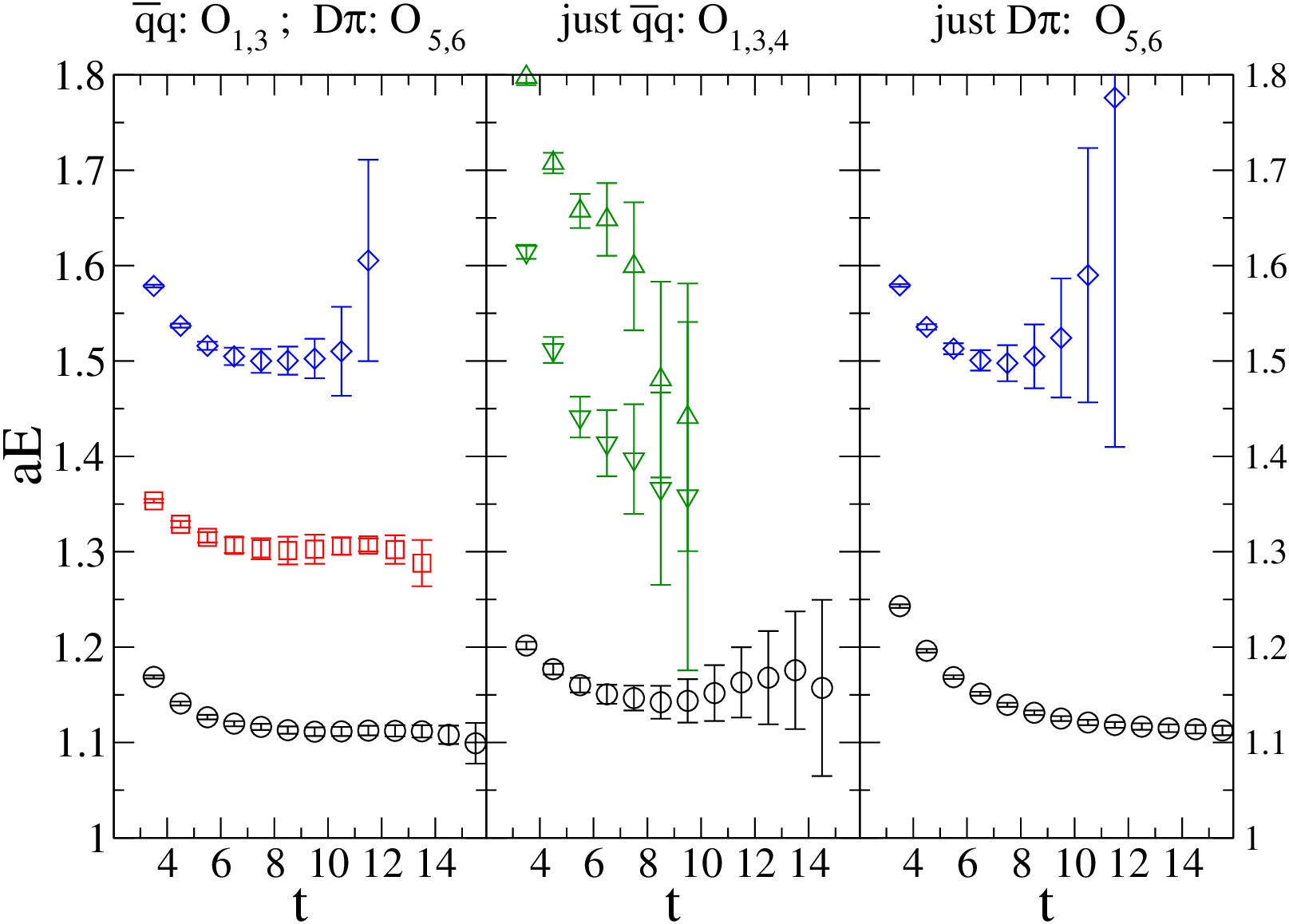}
\caption{Energy levels in the $J^P=0^+$ channel for three different choices of interpolator basis. The panel on the left shows the full results from a basis ${\cal O}_{1,3,5,6}$ of $u\bar{c}$ and $D\pi$ interpolators. The mid panel shows results from just $u\bar{c}$ interpolators (${\cal O}_{1,3,4}$), while the right panel from just $D\pi$ interpolators (${\cal O}_{5,6}$). Interpolators are  listed in Table \ref{interpolators_dmesons} and Eq. (\ref{MM_0+}). All data are for $t_0=3$. }
\label{d0basis}
\end{figure}

\begin{figure}[tb]
\includegraphics[width=85mm,clip]{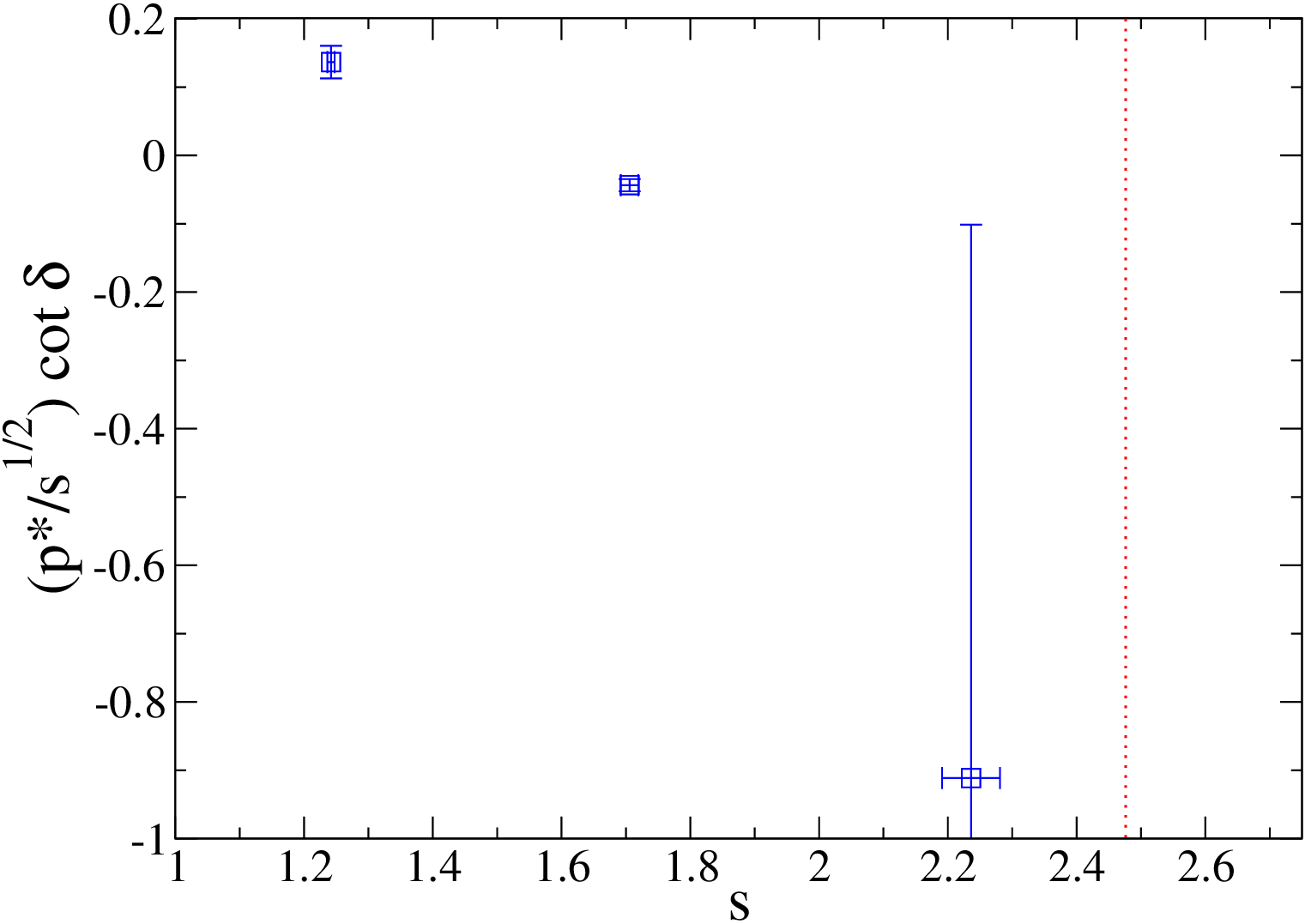}
\caption{The quantity $\rho_0(s)=\frac{p^*}{\sqrt{s}}\cot\delta$  for the $D\pi$ scattering in channel $J^P=0^+$ as a function of $s$, both in units of the lattice spacing. For a single Breit-Wigner type resonance points are expected to lie on a straight line as suggested by (\ref{bwtype},\ref{BW_rho}). The dotted vertical line indicates the $D^*(0)\rho(0)$ threshold with $am_\rho\approx0.51$.}
\label{d0meson_kmatrix}
\end{figure}

We now consider the $J^P=0^+$ channel where only one resonance has been established in experiment \cite{0954-3899-37-7A-075021}, the $D_0^*(2400)$. The first step is to extract the energy levels in the finite volume. Figure \ref{d0meson_effmass} shows the effective masses obtained for the three lowest states in the $D_0^*$ channel for our final selection of interpolating fields. Results from the full basis agree qualitatively but are more noisy. For all displayed states, we can obtain stable fits which are reasonably insensitive to the choice of fit range, number of fit exponentials (one or two) and choice of $t_0$. The inelastic threshold opens at $D^*(0)\rho(0)$, which is at
$E=m_{D^*}+m_{\rho}\approx 1.57$ on our lattice and is situated above $D(1)\pi(-1)$ in Fig. \ref{d0meson_effmass}.

To illustrate the effect of a combined basis consisting of both quark-antiquark and meson-meson interpolators we compare the results for the lowest levels in the $D_0^*$ channel for different choices of interpolator basis in Figure \ref{d0basis}. The results plotted in the left panel correspond to the data from our final choice of interpolators already shown in Figure \ref{d0meson_effmass}. In the right panel we plot data from our $2\times2$ basis of meson-meson interpolators. As expected these interpolators lead to energy levels in the vicinity of the non-interacting $D(0)\pi(0)$ and $D(1)\pi(-1)$ states, but also show sizable excited state contaminations. The mid panel shows results from a basis consisting of a subset of $q\bar{q}$ interpolators. The ground state for this choice turns out to be in the vicinity of the $D(0)\pi(0)$ state but has much larger errors than the ground state from the full basis  in the plateau region. The effective masses calculated from the second and third eigenvalues never plateau and are very noisy. This is quite contrary to the full basis, where the plateau for the level $n=2$ is well determined. From this plot it is quite obvious that an analysis of the energy levels considering only $q\bar{q}$ interpolating fields would not lead to satisfactory results with our sources and statistics.

Table \ref{d0results} shows the results for the preferred interpolator choices that combine $q\bar q$ and meson-meson interpolators and correspond to the levels in Figure \ref{d0meson_effmass}. It provides the momentum $p^*$ defined in (\ref{pstardef}), the invariant mass squared $s$ and the S-wave scattering phase $\delta_0(s)$ extracted using L\"uscher's relation (\ref{luscher_eq}) for the three lowest levels\footnote{$\delta_0$ has been extracted under the assumption that admixtures from higher partial waves due to the broken rotational symmetry on the lattice do not play a significant role. In our case of irrep $A_1^+$ and total momentum $0$, these admixtures enter only at $l\geq 4$ and should be small.}. 
The ground state energy in this attractive channel is below the non-interacting $D(0)\pi(0)$ level and the corresponding $p^*$ and $\delta_0$ are imaginary. While the phases for the first two levels are fairly well determined, our conservative estimate for the third level differs only fairly insignificantly from the non-interacting $D(1)\pi(-1)$ level. As a consistency check, we therefore compared our results for the energy levels with values from the ratio method, used, for example, in the case of the $\rho$ meson by the PACS-CS collaboration \cite{Aoki:2007rd,Aoki:2010hn}. Within errors, the extracted energy levels agree. 

The S-wave $D\pi$ scattering length $a_0=\tan(\delta)/p^*$ (\ref{a0}) is extracted from the lowest level with small $p^*$ 
\begin{align}
\label{a0_Dpi}
a^{I=1/2}_0&=6.56~\pm~ 1.16\, a\nonumber\\
&=0.81~\pm~0.14~\pm~0.01~\mathrm{fm}\;,\\
 \frac{a^{I=1/2}_0}{\mu_{D\pi}}&=17.7~\pm~3.1~\pm~0.2~\mathrm{GeV}^{-2}\;,\nonumber
\end{align}
where $m_{D,\pi}$ from  the simulation were inserted to the reduced mass $\mu_{D\pi}$. The ratio $a_0/\mu_{D\pi}$ is independent of $m_\pi$ within 
 Weinberg's current algebra result $a_0/\mu_{D\pi}=1/(2\pi F_\pi^2)\approx 9.4~$GeV$^{-2}$ with $F_\pi \approx 0.13~$GeV \cite{PhysRevLett.17.616,Liu:2008rza}. Heavy meson ChPT combined with the lattice input from \cite{Liu:2008rza} leads to $a_0/\mu_{D\pi}\approx 9-11~$GeV$^{-2}$ \cite{Liu:2009uz,Geng:2010vw} with physical $m_{D,\pi}$ input to the reduced mass\footnote{Where given we use the masses provided by the authors for calculating the reduced mass. When not provided we use the values from \cite{Guo:2009ct}.}. An indirect determination based on the simulation of the $D\to \pi$ semileptionic transition  leads to $a_0/\mu_{D\pi}=16.4\pm 2.3~$GeV$^{-2}$ \cite{Flynn:2007ki} which agrees with our result within error. A calculation using Unitarized ChPT and taking into account coupled channel effects results in $a_0/\mu_{D\pi}=13.8\pm 0.4~$GeV$^{-2}$ which is also compatible with our result.
It is interesting to note that  our $a_0/\mu\approx 18~$GeV$^{-2}$ for $D\pi$ is very close to our result for $D^* \pi$ (\ref{a0_Dstarpi}) and $K\pi$ \cite{Lang:2012sv} in $I=1/2$ channels. Indeed, current algebra predicts the same ratio for all three channels, albeit the current algebra result itself is lower.   

In Figure \ref{d0meson_kmatrix} we plot $\rho_0(s)=\frac{p^*}{\sqrt{s}}\cot\delta_0$ (\ref{rho}) for the $D_0^*$ channel as a function of $s$. For a single Breit-Wigner type resonance a linear relationship (\ref{BW_rho}) should emerge. Unfortunately this relationship can not be tested with our current data, as our highest energy level is not determined well enough. Assuming a Breit-Wigner amplitude the data for the first two energy levels clearly indicate a resonance between levels 1 and 2. We  extract the resonance mass and coupling $g_{D_0^* D\pi}^{lat}$  from $\rho_0(s)$ with a linear fit (\ref{BW_rho}) over three points\footnote{Due to the large error for $\rho_0(n=3)$, the results are almost independent if the level $n=3$ is taken in the linear fit or not.}. We obtain
\begin{align}
\label{R-D0}
g_{D_0^* D\pi}^{lat}&=2.55(21)(03) \mathrm{GeV}\;,\\
m_{D_0^*}^{lat}-M^{lat}_{\overline{1S}}&=350.8(20.2)(3.7) \mathrm{MeV}\;,
\end{align}
where  $M_{\overline{1S}}=\frac{1}{4}(M_D+3M_{D^*})$. The resulting mass is compared to the PDG value for the $D_0^*(2400)$ in Figure \ref{dmeson_results_channel}. To compare our coupling to experiment, we can use the total width $\Gamma_{D_0^*}=247(40)~$MeV \cite{0954-3899-37-7A-075021} and translate it into an upper bound for the coupling since the branching ratio $\Gamma(D_0^*\to D\pi)/\Gamma_{tot}$ has not been measured (but is expected to be $\simeq 100\%$). The resulting value $g_{D_0^* D\pi}^{exp}\le1.92(14)~$GeV is not too far from our estimate, although slightly smaller. A value of $\Gamma(D_0^*\to D_0\pi)$ close to the experimental one was extracted on the lattice also from an independent method, which is based on the simulation of soft pion emission in the kinematical situation, where the initial and the final heavy mesons are at rest and $m_c\to \infty$ \cite{Becirevic:2012zza}. \footnote{A detailed lattice study of the various couplings for the soft pion emission from the static heavy hadrons (but not $D_0^*$ or $D_1$ was recently also considered in \cite{Detmold:2011bp,Detmold:2012ge}.}

\subsubsection{$D^* \pi$ scattering in the $J^P=1^+$ channel and $D_1$ resonances}

\begin{figure}[tbh]
\includegraphics[width=85mm,clip]{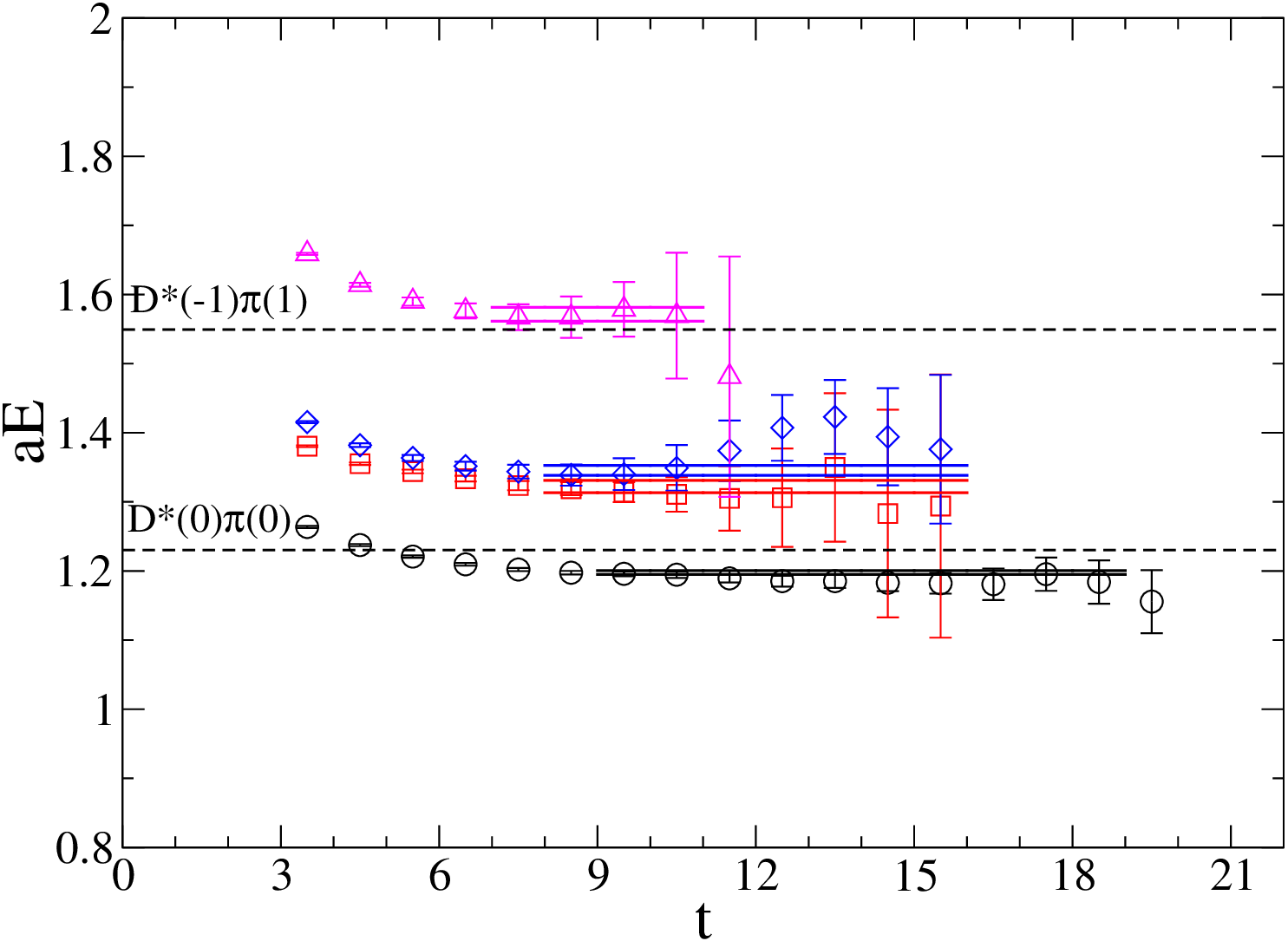}
\caption{The lowest four energy levels in the $D_1^*$ channel with $J^P=1^+$. The fit ranges and uncertainties are indicated by the solid horizontal lines. In addition, non-interacting scattering levels $D^*(p)\pi(-p)$ are depicted by dashed lines. Interpolator choices and numerical values can be found in Table \ref{d1results}.} 
\label{d1meson_effmass}
\end{figure}

\begin{figure}[tb]
\includegraphics[width=85mm,clip]{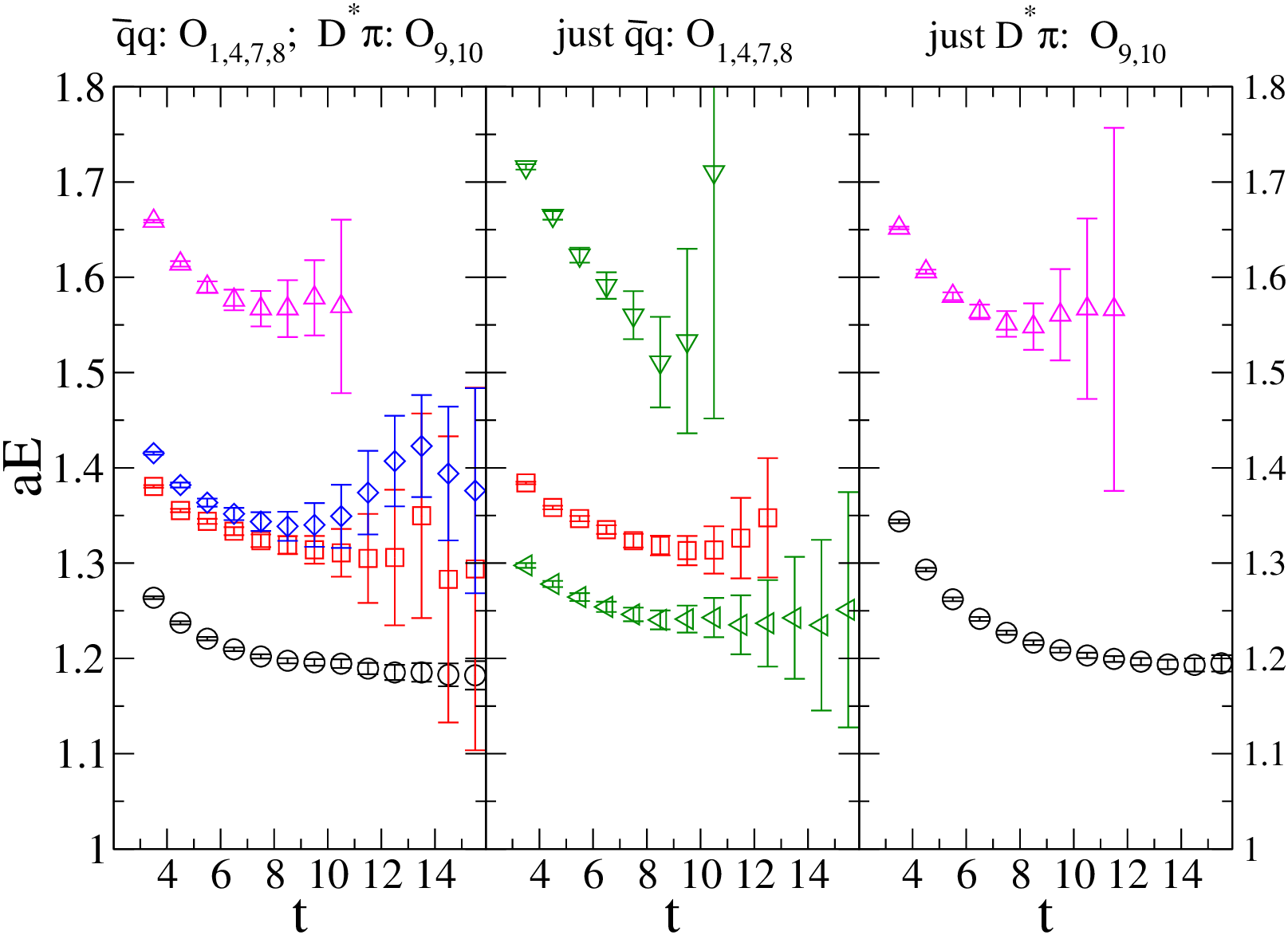}
\caption{Energy levels in the $J^P=1^+$ channel for three different choices of interpolator basis. The panel on the left shows the full results from a basis ${\cal O}_{1,4,7,8,9,10}$ of $u\bar{c}$ and $D^* \pi$ interpolators. The mid panel shows results from just $u\bar{c}$ interpolators (${\cal O}_{1,4,7,8}$), while the right panel contains our results from just $D^* \pi$ interpolators (${\cal O}_{9,10}$). Interpolators are  listed in Table \ref{interpolators_dmesons} and Eq. (\ref{MM_1+}). All data are for $t_0=3$.}
\label{d1basis}
\end{figure}

\begin{figure}[tbh]
\includegraphics[width=85mm,clip]{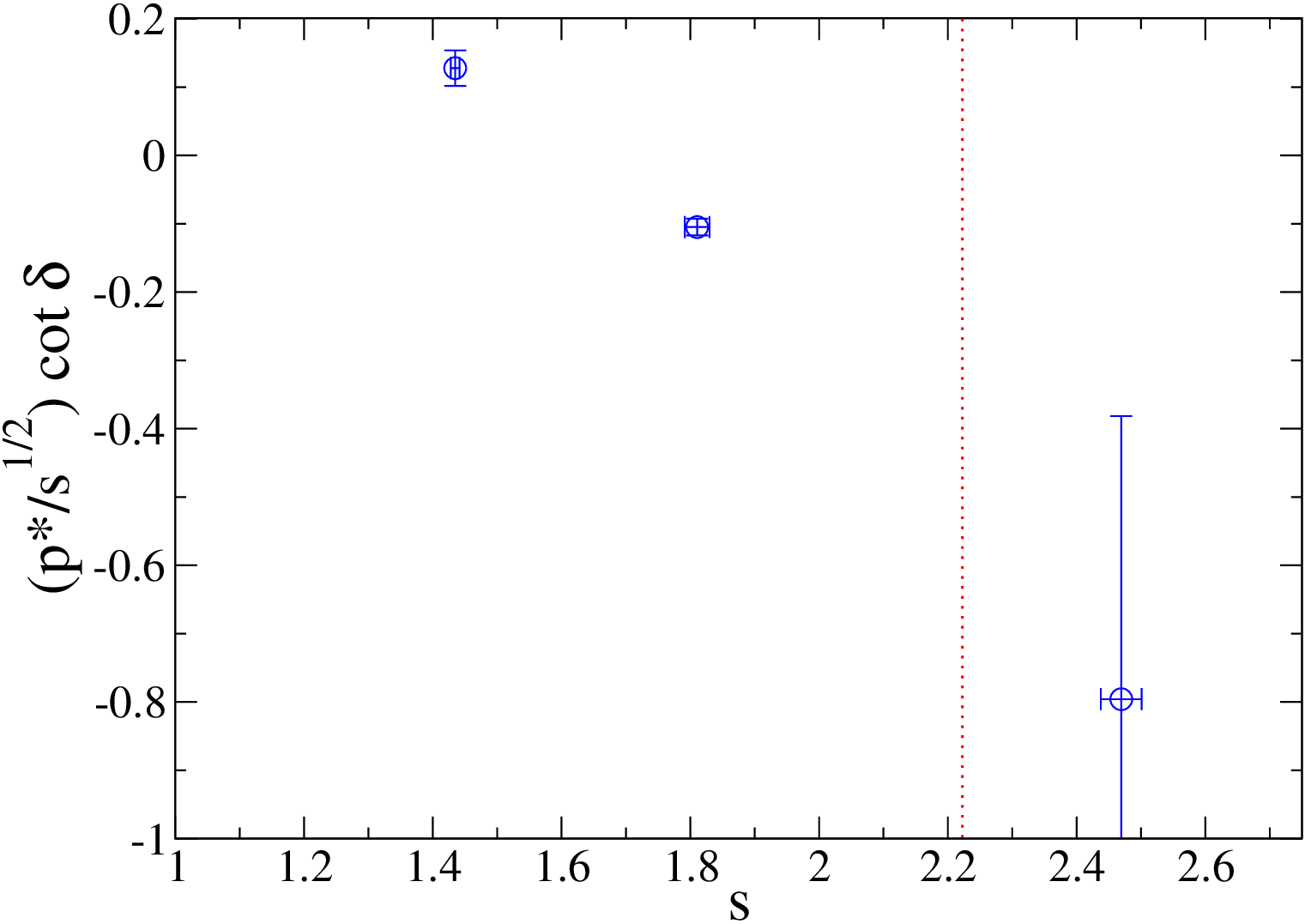}
\caption{Same as Figure \ref{d0meson_kmatrix} for the $D_1$ channel with $J^P=1^+$. The dotted vertical line indicates the $D(0)\rho(0)$ threshold with $am_\rho\approx0.51$.}
\label{d1meson_kmatrix}
\end{figure}

Unlike in the previous case, the $D^*\pi$ scattering with $J^P=1^+$ gets contributions from  S-wave as well as D-wave, and there are two known resonances: the $D_1(2420)$ and the $D_1(2430)$ \cite{0954-3899-37-7A-075021}.  Again, we start by extracting the energy levels for the lowest states from our simulation. The results are plotted in Figure \ref{d1meson_effmass}. From the experience with the $D_0^*$ in the previous section, we expect to extract two energy levels in the vicinity of the lowest scattering states $D^*(0)\pi(0)$ and $D^*(1)\pi(-1)$ and two additional levels related to the two resonances in this channel. As can be seen in Figure \ref{d1meson_effmass} this is the case. 

Again one can compare the results from the final choice of interpolators to subsets containing only $q\bar{q}$ or only meson-meson interpolators. This comparison is shown in Figure \ref{d1basis}. The results from $q\bar{q}$ interpolators alone are shown in the mid panel. The two largest eigenvalues lead to effective masses which seem to display a clear plateau at intermediate source-sink separations. In the right panel the results using only meson-meson interpolators are shown. Here we observe clear signals in the vicinity of the non-interacting $D^*(0)\pi(0)$ and $D^*(1)\pi(-1)$ states. Notice that the lowest level from just $q\bar{q}$ interpolators is at best marginally compatible with the ground state from meson-meson interpolators. Turning our attention to the full basis shown in the left panel, we notice that the ground state is compatible with the state observed from meson-meson interpolators alone, while the $n=2$ level is very similar to the $n=2$ level with just $q\bar{q}$ interpolators. There is no level in the vicinity of the $q\bar{q}$ ground state (green left triangles in the mid panel) but a new state (blue diamonds in the left panel) emerges. The $n=4$ state is found in the vicinity of the non-interacting $D^*(1)\pi(-1)$ level. It is interesting to see that one of the levels observed with just $q\bar{q}$ interpolators survives with no significant change in energy, while the lowest changes quite drastically. We will return to this observation for our interpretation of the data below.

\begin{figure}[tbh]
\includegraphics[width=85mm,clip]{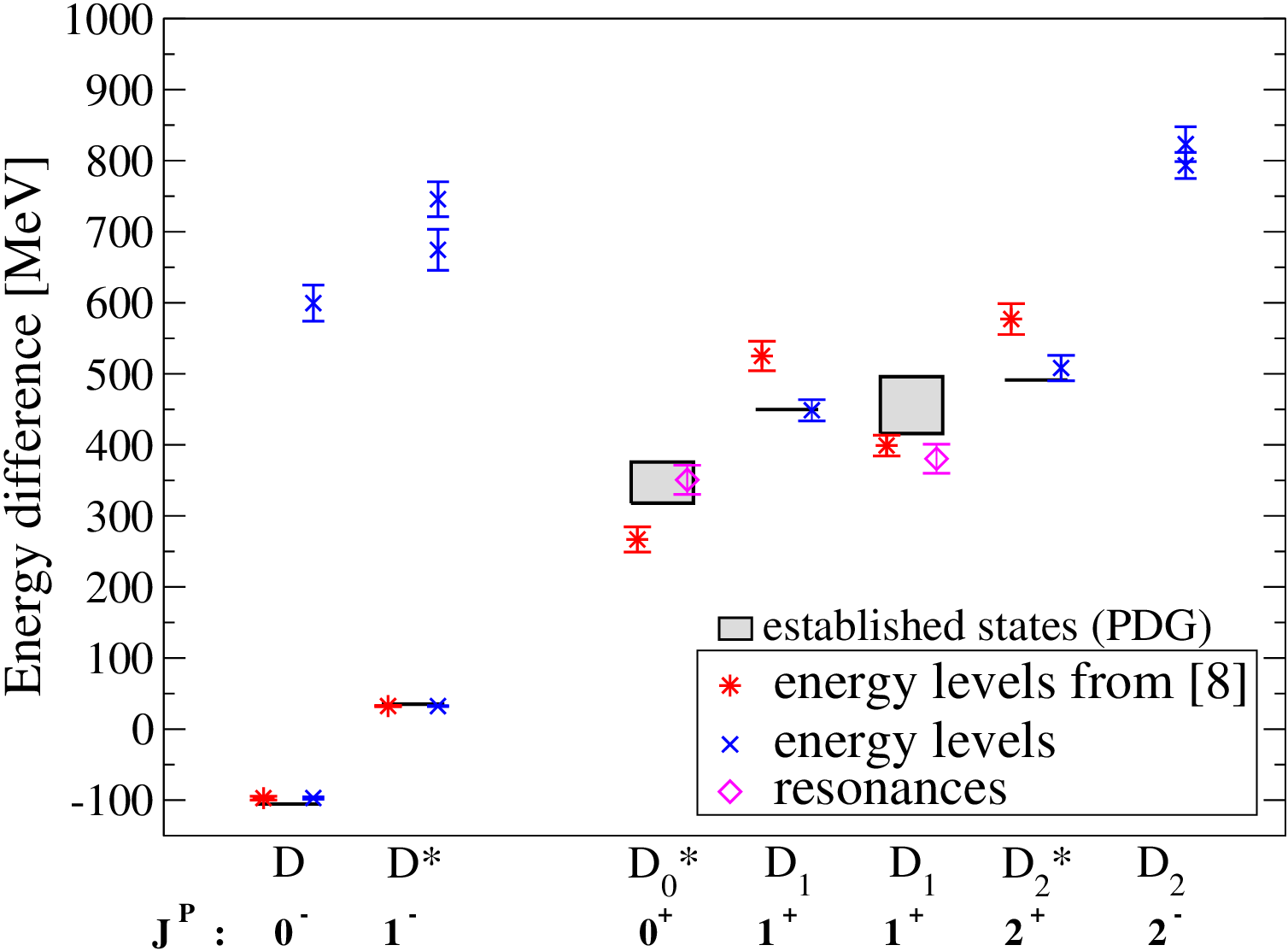}
\caption{Energy differences $\Delta E=E-\tfrac{1}{4}(M_{D}+3M_{D^*})$ for $D$ meson  states on the lattice and in experiment; reference spin-averaged mass is $\tfrac{1}{4}(M_{D}+3M_{D^*})\approx 1971~$MeV in experiment.
 Magenta diamonds give resonance masses for states treated as resonances in the present simulation. Energy levels as extracted in a finite box are given by blue crosses (present simulation) and red stars (simulation \cite{Mohler:2011ke}).  Established experimental states are depicted with black lines or gray boxes with a solid black outline: the height indicates experimental uncertainty in the resonance mass.}
\label{dmeson_results_channel}
\end{figure}

The resulting energy levels that correspond to the final choice of the basis are tabulated in Table \ref{d1results}. 
The extraction of the phase shifts and the resonance parameters from the energy levels is however more challenging  than in the $J^P=0^+$ case since  S-wave and D-wave contribute to $D^*\pi$ scattering with $J^P=1^+$,  and since there are two resonances. A rigorous treatment is not possible with the present data, and we extract information about the two resonances relying on model assumptions. In particular, we  appeal to the knowledge from the $m_c\to \infty$  limit \cite{PhysRevLett.66.1130}, where one expects two $J^P=1^+$ resonances: one broad resonance  with $j^P=\tfrac{1}{2}^+$ which decays into $D^*\pi$ in S-wave, and  one narrow resonance  with $j^P=\tfrac{3}{2}^+$ which only decays to $D^*\pi$ in D-wave. This qualitatively agrees with the experiment, where the $D_1(2430)$ is broad with $\Gamma=384^{+130}_{-110}$ MeV, while the $D_1(2420)$ is fairly narrow with $\Gamma=27.1(2.7)$ MeV \cite{0954-3899-37-7A-075021}. The presence of additional levels in Fig. \ref{d1meson_effmass} is related to resonances, and we will assume that the energy level $E_2$, that is unaffected by the inclusion of meson-meson interpolating fields (red boxes in Figure \ref{d1basis}) corresponds  to the narrow $D_1(2420)$. On our lattice this resonance is expected to be even narrower than in experiment since the phase space for D-wave decay is smaller at $m_\pi\simeq 266~$ MeV, so the resonance mass is very near the energy level and we present the estimate  $m[D_1(2420)]=E_2$ in Table \ref{dmesons_table}   and Figs. \ref{dmeson_results_channel} and \ref{dmeson_results_final}. 

We assume that the contribution of the D-wave scattering is negligible for the other three levels $E_{1,3,4}$, which is a good approximation  for the two levels $E_{1,4}$ away from the sharp D-wave resonance $D_1(2420)$, but represents an approximation for the level $E_{3}$. Under this assumption, the position of the levels $E_{1,3,4}$ depends only on the S-wave phase shifts $\delta_0$  via the L\"uscher relation (\ref{luscher_eq}). So we extract the value $\delta_0(s)$ for each of the three energy levels and present it in Table \ref{d1results}. For the three lowest levels the S-wave phase shift $\delta_0$ is well determined. For the fourth level, which is in the vicinity of the non-interacting $D^*(1)\pi(-1)$ state and has large overlap with interpolator $\mathcal{O}_{10}$, the results have a large uncertainty, just like in the case of the third level in the $J^P=0^+$ channel. The S-wave phase shift $\delta_0$ is dominated by the broad $D_1(2430)$ resonance, so we extract it's mass and $g_{D_1D\pi}$ coupling using a linear Breit-Wigner fit  (\ref{BW_rho})  of $\rho_0(s)=\frac{p^*}{\sqrt{s}}\cot\delta_0$ (\ref{rho})  presented in Figure \ref{d1meson_kmatrix}. Under this assumptions we obtain 
\begin{align}
\label{R-D1}
g_{D_1D\pi}^{lat}&=2.01(15)(02) \mathrm{GeV}\;,\\
m_{D_1}^{lat}-M_{\overline{1S}}^{lat}&=380.7(20.0)(4.0) \mathrm{MeV}\;,
\end{align}
with $M_{\overline{1S}}=\tfrac{1}{4}(M_D+3M_{D^*})$. 
The mass difference with respect to the spin-averaged $1S$ state is plotted in Figure \ref{dmeson_results_channel}. The coupling can be compared to the upper bound $g_{D_1 D\pi}^{exp}\le2.50(40)$, obtained from the experimental total width of the $D_1(2430)$ and taking into account that $\Gamma(D_1\to D^*\pi)/\Gamma_{tot}$ has not been measured (but is expected to be $\simeq100\%$). Considering the assumptions undertaken in this channel as well as the statistic and systematic uncertainties of our simulation both resonance parameters are in reasonable agreement with experiment.

Finally, we  present the S-wave $D^*\pi$ scattering length (\ref{a0})  extracted from the lowest level with small $p^*$ 
\begin{align}
\label{a0_Dstarpi}
a^{I=1/2}_0&=6.53~\pm~1.34\, a\nonumber\\
&=0.81~\pm~0.17~\pm~0.01~\mathrm{fm}\;,\\
 \frac{a^{I=1/2}_0}{\mu_{D\pi}}&=17.6~\pm~3.1~\pm~0.2~\mathrm{GeV}^{-2}\;,\nonumber
\end{align}
which agrees with the result for $D\pi$ (\ref{a0_Dpi}). This can be compared to a calculation using heavy meson ChPT \cite{Liu:2011mi} in which $a_0/\mu_{D\pi}\approx 10.5~$GeV$^{-2}$ is obtained. Just like similar heavy meson ChPT calculations for the $D\pi$ scattering length  \cite{Liu:2009uz,Geng:2010vw,Guo:2009ct} this value is somewhat lower than our result.

\begin{table}[bht]
\begin{center}
\begin{ruledtabular}
\begin{tabular}{|c|c|c|}
 \T\B Mass difference & This paper [MeV] & Experiment [MeV]\\
\hline
\hline
$D_0^*(2400)$ & $350.8\pm20.2\pm3.7$ & $347\pm29$\\
\hline
$D_1(2430)$ & $380.7\pm20.0\pm4.0$ & $456\pm40$\\
\hline
$D_1(2420)$ & $448.7\pm14.1\pm4.7$ & $449.9\pm0.6$ \\
\hline
$D_2^*(2460)$ & $508.2\pm17.1\pm5.3$ & $491.2\pm0.7$ \\
\hline
$D(2550)$ & $599.5\pm24.7\pm6.3$ & $568.0\pm8.2$ \\
\hline
$D^*(2600)$ & $674.5\pm28.0\pm7.1$ & $637.3\pm3.5$ \\
\hline
$D(2750)$ & $793.3\pm16.4\pm8.3$ & $781.0\pm3.2$ \\
\hline
$D^{*\prime}$ & $745.8\pm23.2\pm7.8$ & - \\
\hline
$\overline{2S}-\overline{1S}$ & $655.8\pm24.6\pm6.9$ & $619.9\pm3.7$\\
\hline
1S hyperfine & $129.4\pm1.8\pm1.4$ & $140.65\pm0.10$\\
\hline
2S hyperfine & $75.0\pm26.9\pm0.8$ & $69.3\pm8.9$
\end{tabular}
\end{ruledtabular}
\end{center}
\caption{\label{dmesons_table}Mass differences in the D meson spectrum, compared to experimental values (well established states calculated from \cite{0954-3899-37-7A-075021} and others from \cite{delAmoSanchez:2010vq}).  Bars denote spin-averaged values. For the results of this paper, the first error denotes the statistical uncertainty and the second error denotes the uncertainty from setting the lattice scale. Regarding the scale-setting uncertainty similar remarks to the Charmonium case apply. In addition there is a non-negligible error from the uncertainty in the determination of $\kappa_c$ for all spin-dependent quantities. It should be stressed that the gauge ensembles at our disposal do not allow for a continuum and infinite volume extrapolation. Consequently qualitative but not quantitative agreement is expected.}
\end{table}

\subsubsection{Compilation of $D$ meson results}

Figure \ref{dmeson_results_channel} summarizes our results for the low-lying D meson states compared to well-established experimental results \cite{0954-3899-37-7A-075021} (black lines and boxes). Only the masses (magenta diamonds) of states corresponding to the broad resonances $D_0^*(2400)$ and $D_1(2430)$ have been extracted taking their resonance nature into account explicitly. These are the only two states (among six $1S$ and $1P$ states) that are expected to be broad in the $m_c\to \infty$ limit and are broad in experiment.  The remaining four states are very narrow and can be treated as stable on our lattice, so we equate their masses  to the energy levels determined from correlation functions using only $q\bar{q}$ interpolators. We  make the same assumption for the  states in the $J^P=2^-$ channel and for excited states with $J^P=0^-,~1^-$. 
For the hadronically stable states ($D$, $D^*$ at our simulated pion mass) neglecting explicit coupling to multi-hadron states should be a good approximation. For narrow states above hadronic thresholds one might expect the neglect
of explicit coupling to result in a mass shift comparable to the hadronic width. In addition to the results from this work we also display the results from \cite{Mohler:2011ke} as red stars. In this case, the masses of all states correspond to energy levels determined directly from correlation functions using only $q\bar{q}$ interpolators. As already observed for charmonium, our results for the 1P and 2S states come out at somewhat lower mass than in the previous simulation \cite{Mohler:2011ke}. As we are working with a slightly improved heavy-quark treatment, different sources, a different pion mass, a different volume and a different scale setting procedure it is not clear what exactly causes this difference.  

In addition to the well established states some new resonances were recently observed by the BaBar collaboration \cite{delAmoSanchez:2010vq}. In particular BaBar observes two new resonances\footnote{For all D meson results we always compare to the neutral states.} $D(2550)$ and $D^*(2600)$ which are interpreted in the literature as the 2S states \cite{Sun:2010pg,Li:2010vx,Wang:2010ydc,Zhong:2010vq,Chen:2011rr,Badalian:2011tb}. In addition there is also evidence for a state at a mass of $2752.4\pm1.7\pm2.7~$MeV in $D^*\pi$ and an observation of a state at a mass of $2763.3\pm2.3\pm2.3~$MeV in $D\pi$. While these signals are interpreted as a single state in the PDG \cite{0954-3899-37-7A-075021} and by some authors \cite{Sun:2010pg,Li:2010vx}, others \cite{Wang:2010ydc,Zhong:2010vq,Chen:2011rr,Badalian:2011tb} prefer the interpretation as two different states. If interpreted as two different states, the $D(2750)$ is most commonly interpreted as either the lowest or first excited state in the $2^-$ channel, the $D_2$ or $D_2^\prime$. The $D^*(2760)$ is most commonly interpreted as the ground state in the $J^P=3^-$ channel, the $D^*_3$, although quantum numbers $J^P=1^-$ can not be excluded either, especially if the two observed signals come from the same resonance. 

\begin{figure}[tbh]
\includegraphics[width=85mm,clip]{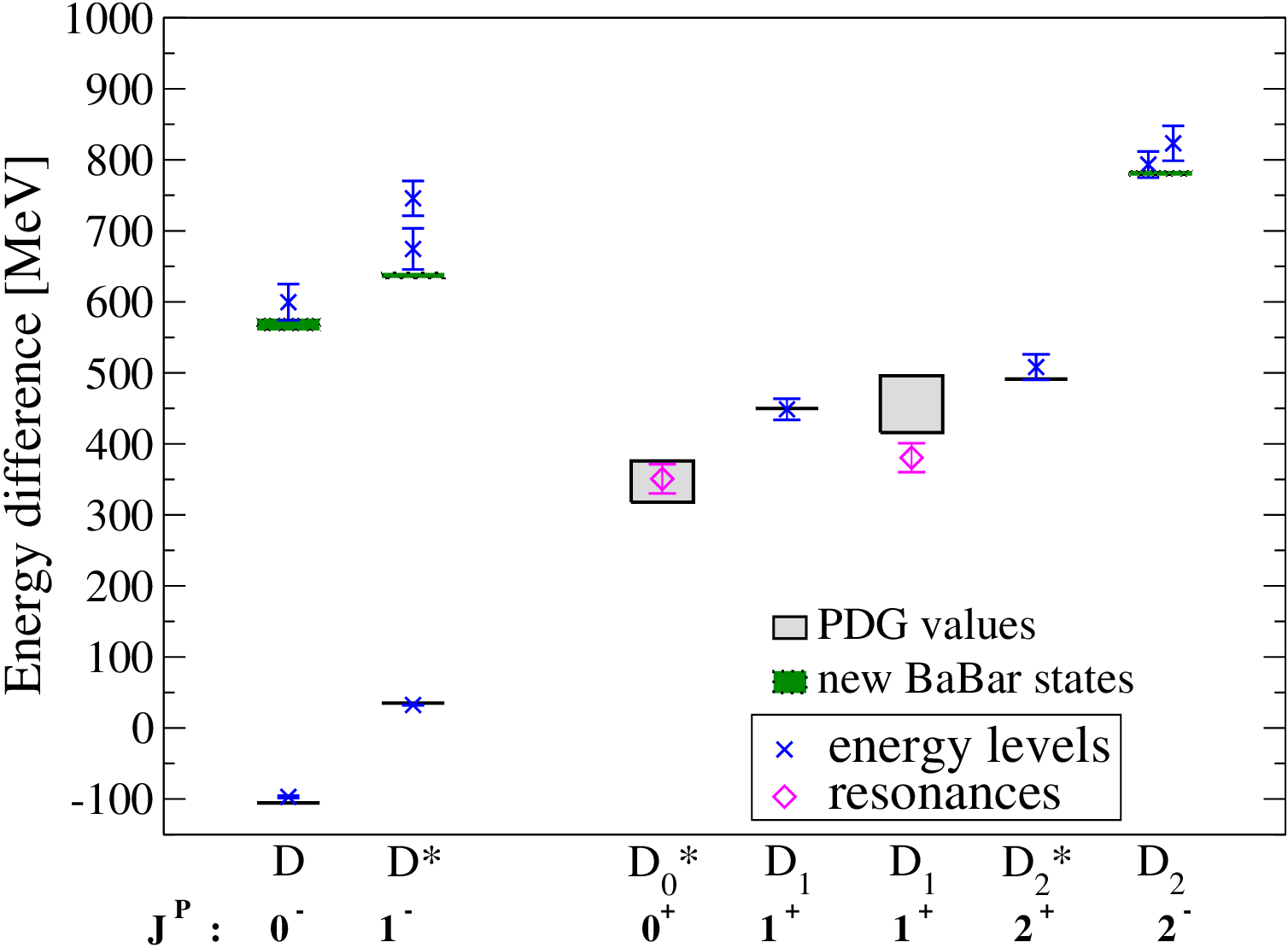}
\caption{Energy differences $\Delta E=E-\tfrac{1}{4}(M_{D}+3M_{D^*})$ for $D$ meson  states in the present simulation  and in experiment; the reference spin-averaged mass is $\tfrac{1}{4}(M_{D}+3M_{D^*})\approx 1971~$MeV in experiment. Magenta diamonds give resonance masses for states treated as resonances in the present simulation. Masses extracted as energy levels in a finite box are displayed as blue crosses. Established experimental states are depicted with black lines or gray boxes with a solid black outline: the height indicates experimental uncertainty in the resonance mass. 
In addition to these well-established states the plot also shows energy levels 
from a recent publication by the BaBar collaboration \cite{delAmoSanchez:2010vq} as green boxes with a dotted black outline, choosing a set of possible quantum number assignments which seems to be favored in the literature \cite{Sun:2010pg,Li:2010vx,Wang:2010ydc,Zhong:2010vq,Chen:2011rr,Badalian:2011tb}. For further comments regarding this assignment please refer to the text.} 
\label{dmeson_results_final}
\end{figure}

In Figure \ref{dmeson_results_final} our results are plotted again with the most likely assignments of these new BaBar states under the assumption of two distinct resonances for the $D(2750)$ and $D^*(2760)$ signals. Table \ref{dmesons_table} provides the numerical values for the mass splittings and also includes our results for the $1S$ and $2S$ hyperfine splittings and for the $\overline{2S}-\overline{1S}$ splitting, where correlations have been taken into account.

\section{\label{outlook}Summary \& discussion}

Charm-light mesons were studied using a dynamical lattice QCD simulation with two flavors of light quarks. It is the first exploratory simulation which  treats  the experimentally broad scalar and axial $D$ mesons as hadronic resonances in $D\pi$ and $D^*\pi$ scattering. A single ensemble with $m_\pi\simeq 266~$MeV and $a\simeq 0.124~$fm is used. A rather small volume $V=16^3\times 32$ enables us to use the costly distillation method, which facilitates the construction of the correlators that incorporate $D\pi$ and $D^*\pi$ operators in addition to the usual quark-antiquark ones.

The heavy quark was treated using the Fermilab approach and the charm quark mass was tuned to fit the kinetic mass of the spin-averaged $1S$ charmonium states. 
As a check, the kinetic masses for spin-averaged S-wave $D$ and $D_s$ mesons were also calculated. At our final choice $\kappa_c=0.123$ the tuned kinetic masses agree with experimental values to better that $2\%$. 

The low-lying charmonium spectrum  was calculated first, to validate our heavy-quark treatment. The distillation method combined with a large basis of quark-antiquark operators allowed the extraction of the ground and a number of excited states. 
The resulting spectrum for various $J^{PC}$ up to $J=3$ in Fig. \ref{charm_results_channel} shows overall good agreement with experiment. An interesting feature of the calculation is the observation of a 
state in the $1^{++}$ channel very close in mass to the $X(3872)$. We however stress that within our systematic uncertainty we can not rule out the possibility of quantum numbers $2^{-+}$ for the X(3872).

Encouraged by that, the S-wave  phase shifts were calculated for $D\pi$ scattering with $J^P=0^+$ and $D^*\pi$ scattering with $J^P=1^+$, focusing on isospin $1/2$ channels where resonances appear. 
 Following the  L\"uscher method, we first extracted the discrete energies of the $D^{(*)}\pi$ system with zero total momentum in a finite box. The energy levels were obtained using quark-antiquark and two-meson operators in the correlation functions. The L\"uscher formula then renders the phase shift for levels in the elastic region. 
 
In the $J^P=0^+$ channel, we extract values of the phase shift at three different relative momenta. 
Only one low-lying resonance is expected in this channel and, assuming a Breit-Wigner shape, a resonance mass and width were extracted. 
The resulting resonance  mass is $351\pm 21~$MeV above the spin-average $\tfrac{1}{4}(m_D+3m_{D^*})$. This agrees with the mass of the observed resonance $D_0^*(2400)$, whose mass is  $347\pm 29~$MeV above the spin average in experiment. We parametrized the  width $\Gamma\equiv g^2p^*/s$, and the resulting $D_0^*\to D\pi$ coupling $g^{lat}=2.55\pm 0.21~$GeV is close to the experimental value $g^{exp}\le1.92\pm 0.14~$GeV.  

The $J^P=1^+$ channel is more complicated due to the presence of two nearby axial resonances. Four energy levels were extracted. One of the levels was essentially unaffected whether the $D^*\pi$ interpolators were included or not, so we associated this level with the narrow $D_1(2420)$. The remaining energy levels were used in a Breit-Wigner fit to obtain resonance parameters which are associated with the broad $D_1(2430)$.
The resonance mass is found at $381 \pm 20~$ MeV above $\tfrac{1}{4}(m_D+3m_{D^*})$, which is slightly lower than the experimental value $456\pm40~$MeV, while the coupling $g^{lat}=2.01\pm 0.15~$GeV agrees with $g^{exp}\le2.50\pm 0.40~$GeV. 

The main results for the $D$ meson spectrum are compiled in Fig. \ref{dmeson_results_final}, where resonance masses for scalar and axial mesons are shown together with our results for other $J^P$. The latter were calculated using just quark-antiquark operators and by equating the masses to the energy levels.  Overall good agreement with experimental values of the well established states was obtained. Furthermore additional energy levels were observed in the vicinity of some of the resonances discovered recently by the BaBar collaboration \cite{delAmoSanchez:2010vq}.

In addition to the resonance parameters, the S-wave scattering lengths $a_0$  were determined from the ground states.  The resulting $a_0^{I=1/2}=0.81\pm0.14~$fm for $D\pi$ and $a_0^{I=1/2}=0.81 \pm 0.17~$fm  for $D^*\pi$ apply for $m_\pi=266~$ MeV in our simulation.  

The experimental observation of new $D$ and $D_s$ meson states in the past decade led to a number of challenges for theory and new ideas emerged. An example is the suggestion that explicit $s\bar s$ content should be included 
in the structure of some charm-light meson states (see for example \cite{Dmitrasinovic:2005gc} and references therein). In this context, it is interesting that the present simulation results in favorable agreement with experiment without the inclusion of strange quark content.

\vspace{0.3cm}

{\it Note added:}

The resonance masses in Eqs. (\ref{R-D0}) and (\ref{R-D1}) correspond to the energies where the scattering phase shift is $90^\circ$. The  corresponding pole locations in the complex energy plane are $E_{D_0^*}^p\!-\!M_{\overline {1S}}  =0.15(3) - i ~0.19(3)~$GeV for the scalar resonance and $E_{D_1}^p\!-\!M_{\overline {1S}} =0.32(4)  - i ~0.08(2)~$GeV for the axial vector resonance. Employing the experimental value of the spin-averaged $D$-meson mass $M_{\overline {1S}}^{exp}=\tfrac{1}{4}(m_D+3m_{D^*})$, these yield the pole positions at  $E_{D_0^*}^p=2.12(3) - i ~0.19(3)~$GeV and $E_{D_1}^p=2.29(3)  - i ~0.08(2)~$GeV.

\vspace{0.5cm}

\acknowledgments
We kindly thank Anna Hasenfratz for providing gauge configurations and Martin L\"uscher for making his DD-HMC software available. We would like to thank B. Blossier, C. DeTar, V. Dmitra\ifmmode \check{s}\else \v{s}\fi{}inovi\ifmmode~\acute{c}\else \'{c}\fi{}, B. Golob, C. B. Lang and  L. Leskovec for helpful discussions. The calculations were performed on computing clusters at TRIUMF, Karl-Franzens Universit\"at Graz and Ljubljana. This work is supported in part by the Natural Sciences and Engineering Research Council of Canada (NSERC) and the Slovenian Research Agency.
\newpage
\begin{appendix}
\section{\label{interpolators}Interpolating fields}

\subsection{\label{charmonium_appendix}Charmonium}

Table \ref{interpolators_charmonium} lists the interpolating fields used in Section \ref{charmonium} to study the low-lying charmonium spectrum.  The symbol $\nabla_i$ is used for a single covariant derivative. For a Laplacian-like structure we use $\Delta=\sum_{i=1}^3\nabla_i\nabla_i$. In addition we also consider structures with a symmetrized second derivative of type $D_i=|\epsilon_{ijk}|\nabla_j\nabla_k$ first proposed in \cite{Liao:2002rj} and previously used in \cite{Dudek:2007wv}. Summation over repeated roman indices is implied and $\epsilon_{ijk}$ denotes the Levi-Civita symbol in three dimensions. For interpolating fields in the E representation some non-trivial Clebsch-Gordan coefficients are needed. They are
\begin{align}
Q_{111}=\frac{1}{\sqrt{2}}\;,\quad Q_{122}=-\frac{1}{\sqrt{2}}\;,\nonumber\\
Q_{211}=-\frac{1}{\sqrt{6}}\;,\quad Q_{222}=-\frac{1}{\sqrt{6}}\;,\quad Q_{233}=\frac{2}{\sqrt{6}}\;,\\
Q^\prime_{111}=-\frac{1}{\sqrt{6}}\;,\quad Q^\prime_{122}=-\frac{1}{\sqrt{6}}\;,\quad Q^\prime_{133}=\frac{2}{\sqrt{6}}\;,\nonumber\\
Q^\prime_{211}=-\frac{1}{\sqrt{2}}\;,\quad Q^\prime_{222}=\frac{1}{\sqrt{2}}\;.\nonumber
\end{align}

For completeness, Table \ref{fit_details} lists the interpolator choice, timeslice $t_0$, fit range and type as well as the fit results and the $\chi^2/$d.o.f for all fits performed to determine energy levels related to charmonium states.

\begin{center}
\LTcapwidth=3.2in
\begin{longtable}{|c|c|c|c|}

\caption[Table of interpolating fields]{Table of $c\bar c$ interpolators  used for charmonium states. They are sorted by irreducible representation of the octahedral group $O_h$ and by quantum numbers $PC$. The reduced lattice symmetry implies an infinite number of continuum spins in each irreducible representation of the octahedral group. For operators, repeated roman indices indicate summation. The quantity $\gamma_t$ denotes the Dirac matrix for the time direction. }\label{interpolators_charmonium}\\[0.3cm]
\hline\hline
\T\B  Lattice & Quantum numbers  & Interpolator & Operator \\
\T\B  irrep &  $J^{PC}$ in irrep & label & \\
\endfirsthead

\multicolumn{4}{c}%
{{\T\B\bfseries \tablename\ \thetable{} -- continued from previous column}} \\
\hline
\hline 
\T\B  Lattice & Quantum numbers  & Interpolator & Operator \\
\T\B  irrep &  $J^{PC}$ in irrep & label & \\
\endhead

\hline \multicolumn{4}{|r|}{{\T\B Continued in next column}} \\ \hline
\endfoot

\hline \hline
\endlastfoot

\hline
\T\B $A1^{-+}$ & $0^{-+}$, $4^{-+}$, $\dots$ & 1 & $\bar{q}\gamma_5q$ \\
\T\B  & & 2 & $\bar{q}\gamma_t\gamma_5q$ \\
\T\B  & & 3 & $\bar{q}\gamma_t\gamma_i\gamma_5\overrightarrow{\nabla_i}q$ \\
\T\B  & & 4 & $\bar{q}\overleftarrow{\nabla_i}\gamma_5\overrightarrow{\nabla_i}q$ \\
\T\B  & & 5 & $\bar{q}\overleftarrow{\nabla_i}\gamma_t\gamma_5\overrightarrow{\nabla_i}q$ \\
\T\B  & & 6 & $\bar{q}\overleftarrow{\Delta}\gamma_5\overrightarrow{\Delta}q$ \\
\T\B  & & 7 & $\bar{q}\overleftarrow{\Delta}\gamma_t\gamma_5\overrightarrow{\Delta}q$ \\
\T\B  & & 8 & $\bar{q}\overleftarrow{\Delta}\gamma_t\gamma_i\gamma_5\overrightarrow{\nabla_i}q$ \\
\hline
\T\B $A1^{++}$ & $0^{++}$, $4^{++}$, $\dots$ & 1 & $\bar{q}q$ \\
\T\B  & & 2 & $\bar{q}\gamma_i\overrightarrow{\nabla_i}q$ \\
\T\B  & & 3 & $\bar{q}\gamma_t\gamma_i\overrightarrow{\nabla_i}q$ \\
\T\B  & & 4 & $\bar{q}\overleftarrow{\nabla_i}\overrightarrow{\nabla_i}q$ \\
\T\B  & & 5 & $\bar{q}\overleftarrow{\Delta}\overrightarrow{\Delta}q$ \\
\T\B  & & 6 & $\bar{q}\overleftarrow{\Delta}\gamma_i\overrightarrow{\nabla_i}q$ \\
\T\B  & & 7 & $\bar{q}\overleftarrow{\Delta}\gamma_t\gamma_i\overrightarrow{\nabla_i}q$ \\
\hline
\T\B $T_1^{--}$ & $1^{--}$, $3^{--}$, $4^{--}$, $\dots$ & 1 & $\bar{q}\gamma_iq$ \\
\T\B & & 2 &  $\bar{q}\gamma_t\gamma_iq$ \\
\T\B & & 3 & $\bar{q}\overrightarrow{\nabla_i}q$\\
\T\B & & 4 &  $\bar{q}\epsilon_{ijk}\gamma_j\gamma_5\overrightarrow{\nabla_k}q$\\
\T\B & & 5 & $\bar{q}\overleftarrow{\nabla_i}\gamma_i\overrightarrow{\nabla_i}q$ \\
\T\B & & 6 &  $\bar{q}\overleftarrow{\nabla_i}\gamma_t\gamma_i\overrightarrow{\nabla_i}q$ \\
\T\B & & 7 & $\bar{q}\overleftarrow{\Delta}\gamma_i\overrightarrow{\Delta}q$ \\
\T\B & & 8 &  $\bar{q}\overleftarrow{\Delta}\gamma_t\gamma_i\overrightarrow{\Delta}q$ \\
\T\B & & 9 & $\bar{q}\overleftarrow{\Delta}\overrightarrow{\nabla_i}q$\\
\T\B & & 10 &  $\bar{q}\overleftarrow{\Delta}\epsilon_{ijk}\gamma_j\gamma_5\overrightarrow{\nabla_k}q$\\
\T\B & & 11 & $\bar{q}|\epsilon_{ijk}|\gamma_j\overrightarrow{D_k}q$\\
\T\B & & 12 & $\bar{q}|\epsilon_{ijk}|\gamma_t\gamma_j\overrightarrow{D_k}q$\\
\T\B & & 13 & $\bar{q}\gamma_5\overrightarrow{B_i}q$ \\
\T\B & & 14 & $\bar{q}\gamma_t\gamma_5\overrightarrow{B_i}q$ \\
\hline
\T\B $T_1^{++}$ & $1^{++}$, $3^{++}$, $4^{++}$, $\dots$ & 1 & $\bar{q}\gamma_i\gamma_5q$ \\
\T\B & & 2 & $\bar{q}\epsilon_{ijk}\gamma_j\overrightarrow{\nabla_k}q$\\
\T\B & & 3 & $\bar{q}\epsilon_{ijk}\gamma_t\gamma_j\overrightarrow{\nabla_k}q$\\
\T\B & & 4 & $\bar{q}\overleftarrow{\nabla_i}\gamma_i\gamma_5\overrightarrow{\nabla_i}q$ \\
\T\B & & 5 & $\bar{q}\overleftarrow{\Delta}\gamma_i\gamma_5\overrightarrow{\Delta}q$ \\
\T\B & & 6 & $\bar{q}\overleftarrow{\Delta}\epsilon_{ijk}\gamma_j\overrightarrow{\nabla_k}q$\\
\T\B & & 7 & $\bar{q}\overleftarrow{\Delta}\epsilon_{ijk}\gamma_t\gamma_j\overrightarrow{\nabla_k}q$\\
\T\B & & 8 & $\bar{q}|\epsilon_{ijk}|\gamma_5\gamma_j\overrightarrow{D_k}q$\\
\hline
\T\B $T_1^{+-}$ & $1^{+-}$, $3^{+-}$, $4^{+-}$, $\dots$& 1 & $\bar{q}\gamma_t\gamma_i\gamma_5q$\\
\T\B & & 2 & $\bar{q}\gamma_5\overrightarrow{\nabla_i}q$\\
\T\B & & 3 & $\bar{q}\gamma_t\gamma_5\overrightarrow{\nabla_i}q$\\
\T\B & & 4 & $\bar{q}\overleftarrow{\nabla_i}\gamma_t\gamma_i\gamma_5\overrightarrow{\nabla_i}q$\\
\T\B & & 5 & $\bar{q}\overleftarrow{\Delta}\gamma_t\gamma_i\gamma_5\overrightarrow{\Delta}q$\\
\T\B & & 6 & $\bar{q}\overleftarrow{\Delta}\gamma_5\overrightarrow{\nabla_i}q$\\
\T\B & & 7 & $\bar{q}\overleftarrow{\Delta}\gamma_t\gamma_5\overrightarrow{\nabla_i}q$\\
\T\B & & 8 & $\bar{q}|\epsilon_{ijk}|\gamma_t\gamma_5\gamma_j\overrightarrow{D_k}q$\\
\hline
\T\B $T_2^{--}$ & $2^{--}$, $3^{--}$, $4^{--}$, $\dots$ & 1 & $\bar{q}_s|\epsilon_{ijk}|\gamma_j\gamma_5q^\prime_{D_k}$\\
\T\B & & 2 & $\bar{q}\overleftarrow{\Delta}|\epsilon_{ijk}|\gamma_j\gamma_5q^\prime_{D_k}$\\
\hline
\T\B $T_2^{-+}$ & $2^{-+}$, $3^{-+}$, $4^{-+}$, $\dots$ & 1 & $\bar{q}|\epsilon_{ijk}|\gamma_t\gamma_j\gamma_5\overrightarrow{\nabla_k}q$\\
\T\B & & 2 & $\bar{q}\overleftarrow{\Delta}|\epsilon_{ijk}|\gamma_t\gamma_j\gamma_5\overrightarrow{\nabla_k}q$\\
\T\B & & 3 & $\bar{q}\gamma_5\overrightarrow{D_i}q$\\
\T\B & & 4 & $\bar{q}\gamma_t\gamma_5\overrightarrow{D_i}q$\\
\hline
\T\B $T_2^{++}$ & $2^{++}$, $3^{++}$, $4^{++}$, $\dots$ & 1 & $\bar{q}|\epsilon_{ijk}|\gamma_j\overrightarrow{\nabla_k}q$\\
\T\B & & 2 & $\bar{q}|\epsilon_{ijk}|\gamma_t\gamma_j\overrightarrow{\nabla_k}q$\\
\T\B & & 3 & $\bar{q}\overleftarrow{\Delta}|\epsilon_{ijk}|\gamma_j\overrightarrow{\nabla_k}q$\\
\T\B & & 4 & $\bar{q}\overleftarrow{\Delta}|\epsilon_{ijk}|\gamma_t\gamma_j\overrightarrow{\nabla_k}q$\\
\T\B & & 5 & $\bar{q}\overrightarrow{D_i}q$\\
\hline
\T\B $E^{--}$ & $2^{--}$, $4^{--}$, $\dots$ & 1 & $\bar{q}Q_{ijk}\gamma_j\gamma_5\overrightarrow{\nabla_k}q$\\
\T\B & & 2 & $\bar{q}\overleftarrow{\Delta}Q_{ijk}\gamma_j\gamma_5\overrightarrow{\nabla_k}q$\\
\T\B & & 3 & $\bar{q}Q^\prime_{ijk}\gamma_j\overrightarrow{D_k}q$\\
\T\B & & 4 & $\bar{q}Q^\prime_{ijk}\gamma_t\gamma_j\overrightarrow{D_k}q$\\
\hline
\T\B $E^{-+}$ & $2^{-+}$, $4^{-+}$, $\dots$ & 1 & $\bar{q}Q_{ijk}\gamma_t\gamma_j\gamma_5\overrightarrow{\nabla_k}q$\\
\T\B & & 2 & $\bar{q}Q_{ijk}\gamma_t\gamma_j\gamma_5\overrightarrow{D_k}q$\\
\hline
\T\B $E^{++}$ & $2^{++}$, $4^{++}$, $\dots$ & 1 & $\bar{q}Q_{ijk}\gamma_j\overrightarrow{\nabla_k}q$\\
\T\B & & 2 & $\bar{q}Q_{ijk}\gamma_t\gamma_j\overrightarrow{\nabla_k}q$\\
\T\B & & 3 & $\bar{q}\overleftarrow{\Delta}Q_{ijk}\gamma_j\overrightarrow{\nabla_k}q$\\
\T\B & & 4 & $\bar{q}\overleftarrow{\Delta}Q_{ijk}\gamma_t\gamma_j\overrightarrow{\nabla_k}q$\\
\T\B & & 5 & $\bar{q}Q^\prime_{ijk}\gamma_5\gamma_j\overrightarrow{D_k}q$\\
\hline
\T\B $A_2^{++}$ & $3^{++}$, $6^{++}$, $\dots$ & 1 & $\bar{q}\gamma_5\gamma_i\overrightarrow{D_i}q$\\
\hline
 \T\B $A_2^{+-}$ & $3^{+-}$, $6^{+-}$, $\dots$ & 1 & $\bar{q}\gamma_t\gamma_5\gamma_i\overrightarrow{D_i}q$\\
\hline
\T\B $A_2^{--}$ & $3^{--}$, $6^{--}$, $\dots$ & 1 & $\bar{q}\gamma_i\overrightarrow{D_i}q$\\
\T\B & & 2 & $\bar{q}\gamma_t\gamma_i\overrightarrow{D_i}q$\\
\end{longtable}
\end{center}

\subsection{\label{fit_met}Fitting methodology}

Depending on the channel the full interpolator basis is pruned to a less noisy subset. We either fit with a single exponential at large time separations or with two exponentials starting at smaller time separations. A jackknife estimate of the covariance matrix on the ensemble average is used to perform correlated fits. To build the pseudo-inverse of the matrix we perform a singular value decomposition and exclude very tiny singular values when the ratio of largest over smallest singular values gets close to machine precision. In these cases it is necessary to remove the corresponding number of degrees of freedom from the fit. Table \ref{fit_details} lists our choices for the interpolator sets, timeslice $t_0$ and fit ranges as well as the fit results and $\chi^2/$d.o.f. for the fits. The basis used is indicated by the interpolator numbers from Table \ref{interpolators_charmonium}.

\begin{table*}[t]
\begin{ruledtabular}
\begin{tabular}{cccccccc}
\T\B channel & state & interpolators & $t_0$ & fit range  & fit type & $E_na$ & $\chi^2/$d.o.f \\ 
\hline
$A1^{-+}$& 1 & 1,3,4,6,8 & 2 & 3-27 & 2 exp & 1.47392(31) & 13.68/7\\
         & 2 & 1,3,4,6,8 & 2 & 3-18 & 2 exp & 1.8694(32) & 15.47/7\\
         & 3 & 1,3,4,6,8 & 2 & 3-11 & 2 exp & 2.145(21) & 1.69/5\\
$A1^{++}$& 1 & 1,2,3,4,6,7 & 3 & 4-21 & 2 exp & 1.7475(29) & 4.57/8 \\
         & 2 & 1,2,3,4,6,7 & 3 & 7-12 & 1 exp & 2.021(13) & 0.63/4 \\
         & 3 & 1,2,3,4,6,7 & 3 & 7-11 & 1 exp & 2.154(31) & 1.10/3 \\
$T1^{--}$& 1 & 1,2,3,4,5,6,7,8,11,12 & 2 & 3-27 & 2 exp & 1.54171(43) & 6.75/6 \\
         & 2 & 1,2,3,4,5,6,7,8,11,12 & 2 & 3-16 & 2 exp & 1.9058(33) & 18.35/7 \\
         & 3 & 1,2,3,4,5,6,7,8,11,12 & 2 & 3-15 & 2 exp & 1.9838(47) & 3.80/7 \\
         & 4 & 1,2,3,4,5,6,7,8,11,12 & 2 & 3-14 & 2 exp & 2.0144(45) & 7.45/8 \\
         & 5 & 1,2,3,4,5,6,7,8,11,12 & 2 & 3-9 & 2 exp & 2.172(17) & 2.28/3 \\
         & 6 & 1,2,3,4,5,6,7,8,11,12 & 2 & 3-9 & 2 exp & 2.189(40) & 0.9/3 \\
$T1^{++}$& 1 & 1,3,4,5,7,8 & 3 & 4-14 & 2 exp & 1.7921(26) & 8.27/7 \\
         & 2 & 1,3,4,5,7,8 & 3 & 8-13 & 1 exp & 2.041(18) & 0.64/4 \\
         & 3 & 1,3,4,5,7,8 & 3 & 7-12 & 1 exp & 2.146(15) & 2.06/4 \\
$T1^{+-}$& 1 & 1,3,4,5,7,8 & 3 & 4-14 & 2 exp & 1.7998(32) & 6.14/7 \\
         & 2 & 1,3,4,5,7,8 & 3 & 8-12 & 1 exp & 2.053(20) & 1.57/3 \\
         & 3 & 1,3,4,5,7,8 & 3 & 7-11 & 1 exp & 2.145(17) & 1.21/3 \\
$T2^{--}$& 1 & 1,2  & 2 & 3-16 & 2 exp & 2.0031(43) & 4.09/8 \\
         & 2 & 1,2  & 2 & 3-14 & 2 exp & 2.063(13) & 5.37/7 \\
$T2^{-+}$& 1 & 1,2,3,4 & 2 & 3-14 & 2 exp & 2.0001(41) & 8.05/7 \\
         & 2 & 1,2,3,4 & 2 & 7-10 & 1 exp & 2.190(56) & 1.01/2 \\
$T2^{++}$& 1 & 1,2,3,5 & 2 & 4-17 & 2 exp & 1.8190(31) & 5.48/7 \\
         & 2 & 1,2,3,5 & 2 & 8-13 & 1 exp & 2.061(25) & 0.76/4 \\
         & 3 & 1,2,3,5 & 2 & 7-12 & 1 exp & 2.143(11) & 1.53/4 \\
$A2^{--}$& 1 & 1,2 & 2 & 3-14 & 2 exp & 2.0124(38) & 2.61/7 \\
$A2^{++}$& 1 & 1 & 2 & 7-12 & 1 exp & 2.164(16) & 1.85/4 \\
$A2^{+-}$& 1 & 1 & 2 & 7-12 & 1 exp & 2.155(20) & 2.04./4 \\
\end{tabular}
\end{ruledtabular}
\caption{\label{fit_details}Fit details for the  charmonium states. Interpolators are listed in Table \ref{interpolators_charmonium}.}
\end{table*}

\subsection{\label{dmesons_appendix}D mesons}

Table \ref{interpolators_dmesons} lists the quark-antiquark interpolating fields used in Section \ref{dmesons}. The notation is the same than for charmonium in the previous section. In addition Table \ref{fit_details_dmesons} lists the interpolator choice, timeslice $t_0$, fit range and type as well as the fit results and the $\chi^2/$d.o.f for those D meson-levels which are either stable under the strong interaction or are narrow and have not been properly treated as resonances in our current study.

\begin{table}[t]
\begin{ruledtabular}
\begin{tabular}{|c|c|c|c|}
\T\B  Lattice & Quantum numbers  & Interpolator & Operator \\
\T\B  irrep &  $J^{PC}$ in irrep & label & \\
\hline
$A1^{-}$ & $0^{-}$, $4^{-}$, $\dots$ & 1 & $\bar{q}\gamma_5q^\prime$ \\
 & & 2 & $\bar{q}\gamma_t\gamma_5q^\prime$ \\
 & & 3 & $\bar{q}\gamma_t\gamma_i\gamma_5\overrightarrow{\nabla_i}q^\prime$ \\
 & & 4 & $\bar{q}\gamma_i\gamma_5\overrightarrow{\nabla_i}q^\prime$ \\
 & & 5 & $\bar{q}\overleftarrow{\nabla_i}\gamma_5\overrightarrow{\nabla_i}q^\prime$ \\
 & & 6 & $\bar{q}\overleftarrow{\nabla_i}\gamma_t\gamma_5\overrightarrow{\nabla_i}q^\prime$ \\
\hline
$A1^{+}$ & $0^{+}$, $4^{+}$, $\dots$ & 1 & $\bar{q}q^\prime$ \\
 & & 2 & $\bar{q}\gamma_i\overrightarrow{\nabla_i}q^\prime$ \\
 & & 3 & $\bar{q}\gamma_t\gamma_i\overrightarrow{\nabla_i}q^\prime$ \\
 & & 4 & $\bar{q}\overleftarrow{\nabla_i}\overrightarrow{\nabla_i}q^\prime$ \\
\hline
 $T_1^{-}$ & $1^{-}$, $3^{-}$, $4^{-}$, $\dots$ & 1 & $\bar{q}\gamma_iq^\prime$ \\
 & & 2 &  $\bar{q}\gamma_t\gamma_iq^\prime$ \\
 & & 3 & $\bar{q}\overrightarrow{\nabla_i}q^\prime$\\
 & & 4 &  $\bar{q}\epsilon_{ijk}\gamma_j\gamma_5\overrightarrow{\nabla_k}q^\prime$\\
 & & 5 & $\bar{q}\gamma_t\overrightarrow{\nabla_i}q^\prime$\\
 & & 6 &  $\bar{q}\epsilon_{ijk}\gamma_t\gamma_j\gamma_5\overrightarrow{\nabla_k}q^\prime$\\
 & & 7 & $\bar{q}\overleftarrow{\nabla_i}\gamma_j\overrightarrow{\nabla_i}q^\prime$ \\
 & & 8 &  $\bar{q}\overleftarrow{\nabla_i}\gamma_t\gamma_j\overrightarrow{\nabla_i}q^\prime$ \\
\hline
 $T_1^{+}$ & $1^{+}$, $3^{+}$, $4^{+}$, $\dots$ & 1 & $\bar{q}\gamma_i\gamma_5q^\prime$ \\
 & & 2 & $\bar{q}\epsilon_{ijk}\gamma_j\overrightarrow{\nabla_k}q^\prime$\\
 & & 3 & $\bar{q}\epsilon_{ijk}\gamma_t\gamma_j\overrightarrow{\nabla_k}q^\prime$\\
 & & 4 & $\bar{q}\gamma_t\gamma_i\gamma_5q^\prime$\\
 & & 5 & $\bar{q}\gamma_5\overrightarrow{\nabla_i}q^\prime$\\
 & & 6 & $\bar{q}\gamma_t\gamma_5\overrightarrow{\nabla_i}q^\prime$\\
 & & 7 & $\bar{q}\overleftarrow{\nabla_i}\gamma_j\gamma_5\overrightarrow{\nabla_i}q^\prime$ \\
 & & 8 & $\bar{q}\overleftarrow{\nabla_i}\gamma_t\gamma_j\gamma_5\overrightarrow{\nabla_i}q^\prime$\\
\hline
$T_2^{-}$ & $2^{-}$, $3^{-}$, $4^{-}$, $\dots$ & 1 & $\bar{q}_s|\epsilon_{ijk}|\gamma_j\gamma_5\overrightarrow{\nabla_k}q^\prime$\\
 &  & 2 & $\bar{q}|\epsilon_{ijk}|\gamma_t\gamma_j\gamma_5\overrightarrow{\nabla_k}q^\prime$\\
\hline
 $T_2^{+}$ & $2^{+}$, $3^{+}$, $4^{+}$, $\dots$ & 1 & $\bar{q}|\epsilon_{ijk}|\gamma_j\overrightarrow{\nabla_k}q^\prime$\\
 & & 2 & $\bar{q}|\epsilon_{ijk}|\gamma_t\gamma_j\overrightarrow{\nabla_k}q^\prime$\\
\end{tabular}
\end{ruledtabular}
\caption{\label{interpolators_dmesons}Table of $u\bar c$ interpolators used for $D$ mesons; in addition we use $D\pi$ (\ref{MM_0+}) and $D^*\pi$ (\ref{MM_1+}) interpolators for irreps $A_1^+$ and $T_1^+$. Interpolators are sorted by irreducible representation of the octahedral group $O_h$ and by the parity  quantum number $P$. The reduced lattice symmetry implies an infinite number of continuum spins in each irreducible representation of $O_h$. For operators, repeated roman indices indicate summation. The quantity $\gamma_t$ denotes the Dirac matrix for the time direction.}
\end{table}

\begin{table*}[t]
\begin{ruledtabular}
\begin{tabular}{cccccccc}
\T\B channel & state & interpolators & $t_0$ & fit range  & fit type & $E_na$ & $\chi^2/$d.o.f \\ 
\hline
$A1^{-}$& 1 & 1,2,5 & 2 & 4-28 & 2 exp & 0.97952(78) & 15.63/14 \\
        & 2 & 1,2,5 & 2 & 8-14 & 1 exp & 1.417(16) & 0.68/5 \\
$T1^{-}$& 1 & 1,2,3,7 & 2 & 4-28 & 2 exp & 1.0608(15) & 12.45/14 \\
        & 2 & 1,2,3,7 & 2 & 8-12 & 1 exp & 1.464(18) & 2.63/3 \\
        & 3 & 1,2,3,7 & 2 & 8-12 & 1 exp & 1.509(15) & 0.56/3 \\
$T2^{-}$& 1 & 1,2 & 5 & 7-11 & 1 exp & 1.539(10) & 0.40/3 \\
        & 2 & 1,2 & 5 & 7-11 & 1 exp & 1.557(15) & 3.10/3 \\
$T2^{+}$& 1 & 1,2 & 3 & 4-15 & 2 exp & 1.360(11) & 4.85/8 \\
\end{tabular}
\end{ruledtabular}
\caption{\label{fit_details_dmesons}Fit details for the $D$-meson channels that incorporate only $u\bar c$ interpolators, listed in Table \ref{interpolators_dmesons}.  }
\end{table*}

\newpage

\end{appendix}


\begin{thebibliography}{10}

\bibitem{pdg12}
Particle Data Group, J.~Beringer {\em et~al.},
\newblock Phys.Rev. {\bf D86}, 010001 (2012).

\bibitem{Abe:2003zm}
Belle Collaboration, K.~Abe {\em et~al.},
\newblock Phys.Rev. {\bf D69}, 112002 (2004), [arXiv:hep-ex/0307021].

\bibitem{Abe:2004sm}
Belle Collaboration, K.~Abe {\em et~al.},
\newblock Phys.Rev.Lett. {\bf 94}, 221805 (2005), [arXiv:hep-ex/0410091].

\bibitem{Link:2003bd}
FOCUS Collaboration, J.~Link {\em et~al.},
\newblock Phys.Lett. {\bf B586}, 11 (2004), [arXiv:hep-ex/0312060].

\bibitem{Aubert:2009wg}
BABAR Collaboration, B.~Aubert {\em et~al.},
\newblock Phys.Rev. {\bf D79}, 112004 (2009), [arXiv:0901.1291].

\bibitem{delAmoSanchez:2010vq}
The BABAR, P.~del Amo~Sanchez {\em et~al.},
\newblock Phys. Rev. {\bf D82}, 111101 (2010), [arXiv:1009.2076].

\bibitem{PhysRevLett.66.1130}
N.~Isgur and M.~B. Wise,
\newblock Phys. Rev. Lett. {\bf 66}, 1130 (1991).

\bibitem{Bigi:2007qp}
I.~Bigi {\em et~al.},
\newblock Eur.Phys.J. {\bf C52}, 975 (2007), [arXiv:0708.1621].

\bibitem{Uraltsev:2004ta}
N.~Uraltsev,
\newblock hep-ph/0406086.

\bibitem{LeYaouanc:2000cj}
A.~Le~Yaouanc {\em et~al.},
\newblock Phys.Lett. {\bf B480}, 119 (2000), [arXiv:hep-ph/0003087].

\bibitem{Blossier:2009vy}
European Twisted Mass Collaboration, B.~Blossier, M.~Wagner and O.~Pene,
\newblock JHEP {\bf 0906}, 022 (2009), [arXiv:0903.2298].

\bibitem{Mohler:2011ke}
D.~Mohler and R.~Woloshyn,
\newblock Phys.Rev. {\bf D84}, 054505 (2011), [arXiv:1103.5506].

\bibitem{Namekawa:2011wt}
PACS-CS Collaboration, Y.~Namekawa {\em et~al.},
\newblock Phys.Rev. {\bf D84}, 074505 (2011), [arXiv:1104.4600].

\bibitem{Bali:2011dc}
G.~Bali {\em et~al.},
\newblock PoS {\bf LATTICE2011}, 135 (2011), [arXiv:1108.6147].

\bibitem{Geng:2010vw}
L.~Geng, N.~Kaiser, J.~Martin-Camalich and W.~Weise,
\newblock Phys.Rev. {\bf D82}, 054022 (2010), [arXiv:1008.0383].

\bibitem{Flynn:2007ki}
J.~M. Flynn and J.~Nieves,
\newblock Phys.Rev. {\bf D75}, 074024 (2007), [arXiv:hep-ph/0703047].

\bibitem{Liu:2008rza}
L.~Liu, H.-W. Lin and K.~Orginos,
\newblock PoS {\bf LATTICE2008}, 112 (2008), [arXiv:0810.5412].

\bibitem{PhysRevLett.17.616}
S.~Weinberg,
\newblock Phys. Rev. Lett. {\bf 17}, 616 (1966).

\bibitem{Liu:2009uz}
Y.-R. Liu, X.~Liu and S.-L. Zhu,
\newblock Phys.Rev. {\bf D79}, 094026 (2009), [arXiv:0904.1770].

\bibitem{Guo:2009ct}
F.-K. Guo, C.~Hanhart and U.-G. Meissner,
\newblock Eur.Phys.J. {\bf A40}, 171 (2009), [arXiv:0901.1597].

\bibitem{Liu:2011mi}
Z.-W. Liu, Y.-R. Liu, X.~Liu and S.-L. Zhu,
\newblock Phys.Rev. {\bf D84}, 034002 (2011), [arXiv:1104.2726].

\bibitem{ElKhadra:1996mp}
A.~X. El-Khadra, A.~S. Kronfeld and P.~B. Mackenzie,
\newblock Phys. Rev. {\bf D55}, 3933 (1997), [arXiv:hep-lat/9604004].

\bibitem{Becirevic:2012zza}
D.~Becirevic, E.~Chang and A.~L. Yaouanc,
\newblock 1203.0167.

\bibitem{Detmold:2011bp}
W.~Detmold, C.-J.~D. Lin and S.~Meinel,
\newblock Phys.Rev.Lett. {\bf 108}, 172003 (2012), [arXiv:1109.2480].

\bibitem{Detmold:2012ge}
W.~Detmold, C.~D. Lin and S.~Meinel,
\newblock Phys.Rev. {\bf D85}, 114508 (2012), [arXiv:1203.3378].

\bibitem{Hasenfratz:2008ce}
A.~Hasenfratz, R.~Hoffmann and S.~Schaefer,
\newblock Phys. Rev. D {\bf 78}, 05451 (2008), [arXiv:0806.4586].

\bibitem{Hasenfratz:2008fg}
A.~Hasenfratz, R.~Hoffmann and S.~Schaefer,
\newblock Phys. Rev. D {\bf 78}, 014515 (2008), [arXiv:0805.2369].

\bibitem{Hasenfratz:2007rf}
A.~Hasenfratz, R.~Hoffmann and S.~Schaefer,
\newblock JHEP {\bf 05}, 029 (2007), [arXiv:hep-lat/0702028].

\bibitem{Lang:2011mn}
C.~Lang, D.~Mohler, S.~Prelovsek and M.~Vidmar,
\newblock Phys.Rev. {\bf D84}, 054503 (2011), [arXiv:1105.5636].

\bibitem{Lang:2012sv}
C.~Lang, L.~Leskovec, D.~Mohler and S.~Prelovsek,
\newblock 1207.3204.

\bibitem{Luscher:2007se}
M.~L{\"u}scher,
\newblock JHEP {\bf 07}, 081 (2007), [arXiv:0706.2298].

\bibitem{Luscher:2007es}
M.~L{\"u}scher,
\newblock JHEP {\bf 12}, 011 (2007), [arXiv:0710.5417].

\bibitem{Sasaki:2001nf}
S.~Sasaki, T.~Blum and S.~Ohta,
\newblock Phys. Rev. D {\bf 65}, 074503 (2002), [arXiv:hep-lat/0102010].

\bibitem{Detmold:2008yn}
W.~Detmold, K.~Orginos, M.~J. Savage and A.~Walker-Loud,
\newblock Phys. Rev. D {\bf 78}, 054514 (2008), [arXiv:0807.1856].

\bibitem{Peardon:2009gh}
Hadron Spectrum Collaboration, M.~Peardon {\em et~al.},
\newblock Phys. Rev. D {\bf 80}, 054506 (2009), [arXiv:0905.2160].

\bibitem{Stathopoulos:2009:PPI}
A.~Stathopoulos and J.~R. McCombs,
\newblock {ACM} Transactions on Mathematical Software {\bf 37}, 21:1 (2010).

\bibitem{Oktay:2008ex}
M.~B. Oktay and A.~S. Kronfeld,
\newblock Phys. Rev. {\bf D78}, 014504 (2008), [arXiv:0803.0523].

\bibitem{Burch:2009az}
T.~Burch {\em et~al.},
\newblock Phys. Rev. {\bf D81}, 034508 (2010), [arXiv:0912.2701].

\bibitem{Bernard:2010fr}
Fermilab Lattice Collaboration, MILC Collaboration, C.~Bernard {\em et~al.},
\newblock Phys.Rev. {\bf D83}, 034503 (2011), [arXiv:1003.1937].

\bibitem{0954-3899-37-7A-075021}
Particle Data Group, K.~Nakamura {\em et~al.},
\newblock Journal of Physics G: Nuclear and Particle Physics {\bf 37}, 075021
  (2010).

\bibitem{Luscher:1990ck}
M.~L{\"u}scher and U.~Wolff,
\newblock Nucl. Phys. {\bf B339}, 222 (1990).

\bibitem{Michael:1985ne}
C.~Michael,
\newblock Nucl. Phys. {\bf B259}, 58 (1985).

\bibitem{Blossier:2009kd}
B.~Blossier, M.~Della~Morte, G.~von Hippel, T.~Mendes and R.~Sommer,
\newblock JHEP {\bf 04}, 094 (2009), [arXiv:0902.1265].

\bibitem{Bali:2011rd}
G.~S. Bali, S.~Collins and C.~Ehmann,
\newblock Phys.Rev. {\bf D84}, 094506 (2011), [arXiv:1110.2381].

\bibitem{Liu:2012ze}
Hadron Spectrum Collaboration, L.~Liu {\em et~al.},
\newblock JHEP {\bf 1207}, 126 (2012), [arXiv:1204.5425].

\bibitem{delAmoSanchez:2010jr}
BABAR Collaboration, P.~del Amo~Sanchez {\em et~al.},
\newblock Phys.Rev. {\bf D82}, 011101 (2010), [arXiv:1005.5190].

\bibitem{Choi:2011fc}
S.-K. Choi {\em et~al.},
\newblock Phys.Rev. {\bf D84}, 052004 (2011), [arXiv:1107.0163].

\bibitem{Dudek:2009qf}
J.~J. Dudek, R.~G. Edwards, M.~J. Peardon, D.~G. Richards and C.~E. Thomas,
\newblock Phys. Rev. Lett. {\bf 103}, 262001 (2009), [arXiv:0909.0200].

\bibitem{Dudek:2010wm}
J.~J. Dudek, R.~G. Edwards, M.~J. Peardon, D.~G. Richards and C.~E. Thomas,
\newblock Phys. Rev. {\bf D82}, 034508 (2010), [arXiv:1004.4930].

\bibitem{Lees:2012xs}
BABAR Collaboration, J.~Lees {\em et~al.},
\newblock 1207.2651.

\bibitem{Guo:2012tv}
F.-K. Guo and U.-G. Meissner,
\newblock 1208.1134.

\bibitem{talk:belle}
Belle Collaboration, V.~Bhardwaj,
\newblock Studies of radiative x(3872) decays, 2012,
\newblock Talk at Charm2012
  \url{http://indico.phys.hawaii.edu/getFile.py/access?contribId=21&sessionId=%
48&resId=0&materialId=slides&confId=338}.

\bibitem{Yang:2012my}
Y.-B. Yang {\em et~al.},
\newblock 1206.2086.

\bibitem{Artoisenet:2010va}
P.~Artoisenet, E.~Braaten and D.~Kang,
\newblock Phys.Rev. {\bf D82}, 014013 (2010), [arXiv:1005.2167].

\bibitem{BESIII:2011ab}
BESIII Collaboration,
\newblock Phys.Rev.Lett. {\bf 108}, 222002 (2012), [arXiv:1111.0398].

\bibitem{Zhang:2012tj}
Belle Collaboration, C.~Zhang {\em et~al.},
\newblock 1206.5087.

\bibitem{Becirevic:2012dc}
D.~Becirevic and F.~Sanfilippo,
\newblock 1206.1445.

\bibitem{Briceno:2012wt}
R.~A. Briceno, H.-W. Lin and D.~R. Bolton,
\newblock 1207.3536.

\bibitem{Donald:2012ga}
G.~Donald {\em et~al.},
\newblock 1208.2855.

\bibitem{Olsen:2012xn}
BESIII Collaboration, S.~L. Olsen,
\newblock 1203.4297,
\newblock 20 pages, 15 figures/ an invited talk at the 50th International
  Winter Meeting on Nuclear Physics-Bormio2012, 23-27, January, Bormio, Italy.

\bibitem{Follana:2006rc}
HPQCD Collaboration, UKQCD Collaboration, E.~Follana {\em et~al.},
\newblock Phys.Rev. {\bf D75}, 054502 (2007), [arXiv:hep-lat/0610092].

\bibitem{Luscher:1985dn}
M.~L{\"u}scher,
\newblock Commun. Math. Phys. {\bf 104}, 177 (1986).

\bibitem{Luscher:1986pf}
M.~L{\"u}scher,
\newblock Commun. Math. Phys. {\bf 105}, 153 (1986).

\bibitem{Luscher:1990ux}
M.~L{\"u}scher,
\newblock Nucl. Phys. {\bf B354}, 531 (1991).

\bibitem{Luscher:1991cf}
M.~L{\"u}scher,
\newblock Nucl. Phys. B {\bf 364}, 237 (1991).

\bibitem{Fu:2011xz}
Z.~Fu,
\newblock Phys.Rev. {\bf D85}, 014506 (2012), [arXiv:1110.0319].

\bibitem{Leskovec:2012gb}
L.~Leskovec and S.~Prelovsek,
\newblock Phys.Rev. {\bf D85}, 114507 (2012), [arXiv:1202.2145].

\bibitem{Engel:2011aa}
G.~P. Engel, C.~Lang, M.~Limmer, D.~Mohler and A.~Sch{\"a}fer,
\newblock Phys.Rev. {\bf D85}, 034508 (2012), [arXiv:1112.1601].

\bibitem{Aoki:2007rd}
CP-PACS, S.~Aoki {\em et~al.},
\newblock Phys. Rev. {\bf D76}, 094506 (2007), [arXiv:0708.3705].

\bibitem{Aoki:2010hn}
CS, P.-.~S. Aoki {\em et~al.},
\newblock PoS {\bf LATTICE2010}, 108 (2010), [arXiv:1011.1063].

\bibitem{Sun:2010pg}
Z.-F. Sun, J.-S. Yu, X.~Liu and T.~Matsuki,
\newblock Phys. Rev. {\bf D82}, 111501 (2010), [arXiv:1008.3120].

\bibitem{Li:2010vx}
D.-M. Li, P.-F. Ji and B.~Ma,
\newblock Eur. Phys. J. {\bf C71}, 1582 (2011), [arXiv:1011.1548].

\bibitem{Wang:2010ydc}
Z.-G. Wang,
\newblock Phys. Rev. {\bf D83}, 014009 (2011), [arXiv:1009.3605].

\bibitem{Zhong:2010vq}
X.-H. Zhong,
\newblock Phys. Rev. {\bf D82}, 114014 (2010), [arXiv:1009.0359].

\bibitem{Chen:2011rr}
B.~Chen, L.~Yuan and A.~Zhang,
\newblock Phys.Rev. {\bf D83}, 114025 (2011), [arXiv:1102.4142].

\bibitem{Badalian:2011tb}
A.~Badalian and B.~Bakker,
\newblock Phys.Rev. {\bf D84}, 034006 (2011), [arXiv:1104.1918].

\bibitem{Dmitrasinovic:2005gc}
V.~Dmitra\ifmmode \check{s}\else \v{s}\fi{}inovi\ifmmode~\acute{c}\else
  \'{c}\fi{},
\newblock Phys.Rev.Lett. {\bf 94}, 162002 (2005).

\bibitem{Liao:2002rj}
X.~Liao and T.~Manke,
\newblock hep-lat/0210030.

\bibitem{Dudek:2007wv}
J.~J. Dudek, R.~G. Edwards, N.~Mathur and D.~G. Richards,
\newblock Phys. Rev. {\bf D77}, 034501 (2008), [arXiv:0707.4162].

\end{thebibliography}

\end{document}